\documentclass[useAMS,usenatbib]{mn2e}

\newcommand{\FIG}[1]{}

\usepackage{amssymb}
\usepackage{amsmath}
\usepackage{subfigure}
\usepackage{textcomp}
\usepackage{graphicx}

\title{The formation of protostellar binaries in primordial minihalos}

\author[R. Riaz, S. Bovino, S. Vanaverbeke and  D.R.G. Schleicher ]{R. Riaz$^{1}$\thanks{E-mail: rriaz@astro-udec.cl}
S. Bovino$^{1,2}$\thanks{E-mail: stefano.bovino@uni-hamburg.de} 
S. Vanaverbeke$^{3}$\thanks{E-mail: siegfriedvanaverbeke@gmail.com} D.R.G. Schleicher$^{1}$\thanks{E-mail: dschleicher@astro-udec.cl}\\
$^{1}$Departamento de Astronom\'ia, Facultad Ciencias F\'isicas y Matem\'aticas, Universidad de Concepci\'on, Av. Esteban Iturra s/n Barrio \\
Universitario, Casilla $160$-C, Concepci\'on, Chile \\
$^{2}$Hamburger Sternwarte, Universit\"at Hamburg, Gojenbergsweg 112, 21029 Hamburg, Germany \\
$^{3}$Centre for mathematical Plasma-Astrophysics, Department of Mathematics, KU Leuven, Celestijnenlaan 200B, 3001 Heverlee, Belgium \\ 
}

\begin{document}

\date{Accepted - Received -}

\pagerange{\pageref{firstpage}--\pageref{lastpage}} \pubyear{2015}

\maketitle

\label{firstpage}


\begin{abstract}
The first stars are known to form in primordial gas, either in minihalos with about $10^6$~M$_\odot$ or so-called atomic cooling halos of about $10^8$~M$_\odot$. Simulations have shown that gravitational collapse and disk formation in primordial gas yield dense stellar clusters. In this paper, we focus particularly on the formation of protostellar binary systems, and aim to quantify their properties during the early stage of their evolution. For this purpose, we combine the smoothed particle hydrodynamics code GRADSPH with the astrochemistry package KROME. The GRADSPH-KROME framework is employed to investigate the collapse of primordial clouds in the high-density regime, exploring the fragmentation process and the formation of binary systems. We observe a strong dependence of fragmentation on the strength of the turbulent Mach number $\mathcal{M}$ and the rotational support parameter $\beta{}$. Rotating clouds show significant fragmentation, and have produced several Pop.~III proto-binary systems. We report maximum and minimum mass accretion rates of $2.31 \times 10^{-1}$~M$_{\odot}$ yr$^{-1}$ and $2.18\times 10^{-4}$~M$_{\odot}$ yr$^{-1}$. The mass spectrum of the individual Pop III proto-binary components ranges from $0.88$~M$_{\odot}$ to $31.96$~M$_{\odot}$ and has a sensitive dependence on the Mach number $\mathcal{M}$ as well as on the rotational parameter $\beta{}$. We also report a range from $\sim0.01$ to $\sim1$ for the mass ratio of our proto-binary systems. 
\end{abstract}


\section{Introduction}

It is now well-established that Population III (Pop.~III) stars have formed from primordial gas, where the only available cooling agent in the beginning was molecular hydrogen \citep{Saslaw67, b1, Galli98}. The typical formation sites of the first stars are the so-called minihalos with about $10^6$~M$_\odot$ \citep{b29, b4, b3} or the more massive atomic cooling halos with $\sim10^8$~M$_\odot$, as long as they remained in the primordial state \citep[e.g.][]{Greif08, Wise08}. The cooling in the primordial gas can lower the gas temperature to a minimum of about $300$~K, thereby implying a value for the Jeans mass that is considerably higher than for present-day gas clouds \citep{b4, Glover05}, and requiring an initial dark matter halo mass of at least $10^5$~M$_\odot$ for gravitational collapse \citep{b5}. The protostars must have accreted substantially from their environment in order to achieve high masses \citep{b6,b7,b31}, as the accretion rate is expected to scale with the sound speed $c_s$ to the third power.

In what has become the standard picture, Population III stars are generally considered to be more massive than the current stellar population \citep{b3}. Due to their high masses, temperatures and luminosities, they may have played an important role in shaping the early Universe through radiative and supernova feedback \citep{b8,b9,b10}. The influence of their radiation may have affected the formation of the subsequent generations of stars \citep{Yoshida07, Clark11, Latif14}.  Attempts to understand primordial star formation have started in the late sixth decade of the last century \citep{b11} and continued until more recent investigations \citep{b13,b14, b15}. Recent numerical models based on the statistics of 100 simulations, although partly using 2D radiation hydrodynamical calculations, suggest that the  Initial Mass Function (IMF) has a wide mass distribution $M_{Pop.III}$ = 10 $\sim$ $1000~M_\odot$ \citep{Hirano14}. The IMF for Pop III star formation is therefore still not strongly constrained by current numerical models and needs further investigation \citep{Zackrisson11, Karlsson11}.

A new set of simulations introduced recently \citep{b16,b17,b18,b19} as well as more effective tools to include chemistry into numerical simulations \citep[e.g.][]{b20, Smith05}, have cast serious doubts on the previous understanding related to the morphology of the first generation of stars. While originally Pop.~III stars were assumed to form in isolation \citep{b29,b4,b3}, subsequent simulations going beyond the formation of the first protostars have shown that additional fragments form after the formation of a self-gravitating accretion disk \citep{b2,b22,b23,b15}. This process also leads to the formation of binaries \citep{b24,b25}. Through stellar archeology, indirect constraints are now available on the masses of the first stars which are based upon the chemical abundances observed in extremely metal poor stars \citep[e.g.][]{b26,b27, Frebel15}. These abundance patterns generically suggest that Pop.~III stars exploded as Type II supernovae with progenitors of $15-40$~M$_\odot$ \citep{Umeda03, Keller14, Cooke14}, whereas abundance patterns pointing towards pair-instability supernovae have not been found. A good review on the nucleosynthesis in the first stars was recently provided by \citet{Karlsson13}. 

A change in the mass and multiplicity of Pop.~III stars, as indicated by observations and stellar archeology, has relevant implications for the lifetime of Pop.~III stars, their radiative feedback as well as metal enrichment \citep[e.g.][]{Heger01, Karlsson13}. Additional implications of the formation of Pop.~III binaries include the formation of luminous X-ray binaries \citep{Ricotti16}, which could contribute to the build up of an X-ray background \citep{Glover03}. In addition, the presence of massive close binaries evolving into binary black holes could ultimately lead to detectable emission of gravitational waves due to the merger of massive black holes \citep{Bel17,Miya17,Pacucci17,Schneider17}.

Another interesting result comes from the observation of the Caffau star \citep{Caffau11}. This is the most metal poor star discovered to date. As a result of its extremely low metallicity, in particular the abundances of carbon and oxygen, metal line cooling can be excluded as a relevant cooling mechanism during the formation of this star. These observations therefore point towards fragmentation via dust cooling as a relevant mechanism for the formation of the Caffau star \citep{Schneider12, Klessen12, Bovino16}. To form low-mass stars in a close-to-primordial gas, other mechanisms, such as the ejection of lower-mass objects during three-body interactions, could be a viable formation mechanism \citep{b32,b33}. Ejected Pop.~III stars could stop accreting gas from their surroundings and therefore extend the Pop.~III mass spectrum towards the low-mass end. 

Although a detailed assessment of the Pop.~III mass distribution and the expected binary properties is still out of reach because of computational limitations, our goal for this paper is to assess the properties of protostellar binaries at an intermediate stage of their formation while accretion is still going on. Our simulation results therefore do not correspond to the end product of star formation, but should instead be considered as intermediate results which follow the buildup of the protostars in their early formation stages. This is comparable to the work by \citet{b24}, who evolved a protostellar system for 5000 years after the formation of the first sink particle. Stacy and Bromm have pursued similar simulations for a set of different minihalos. In this paper, we follow a complementary approach by adopting a set of well-defined initial conditions to explore the effect of the initial rotation and turbulence on the outcome of the collapse. Furthermore, we adopt a threshold density of $1.0\times10^{-11}$~g~cm$^{-3}$ for sink formation, an order of magnitude higher as in \citet{b24}, while evolving the resulting system for a similar timescale. The time to evolve the system is defined here through the mass that is accreted onto the protostellar system, and we explore in section 4.8 on how the results depend on the time scale. 

This paper is structured as follows. Section 2 describes the computational scheme employed in our simulations. In section 3 we describe the initial conditions and the setup for our models along with a brief explanation of the treatment of sink particles representing self-gravitating clumps. Section 4 contains a detailed description of our results and our conclusions are presented in section 5.

\section{Computational scheme}

In this paper, we introduce for the first time the coupling of the smoothed Particle Hydrodynamics (SPH) code GRADSPH\footnote{Webpage GRADSPH: http://www.swmath.org/software/1046} \citep{b122} with the chemistry package KROME\footnote{Webpage KROME: http://www.kromepackage.org/} \citep{b20}, which we jointly refer to as GRADSPH-KROME. The combined code allows us to simulate the hydrodynamics of the star forming gas including chemistry and cooling.

In SPH, the density $\rho_{i}$ at the position $\vec{r}_{i}$ of a particle with mass $m_{i}$ is determined by summing the contributions from its neighbours using a weighting function 
$W\left(\vec{r}_{i}-\vec{r}_{j},h_{i}\right)$ with smoothing length $h_{i}$:

\begin{equation} \label{density}
\rho _{i} =\sum _{j} m_{j} W\left(\vec{r}_{i}-\vec{r}_{j},h_{i}\right).
\end{equation}
GRADSPH uses the standard cubic spline kernel with compact support within a smoothing sphere of size $2 h_{i}$ \citep{b123}. The smoothing length $h_{i}$ itself is determined using the following  relation: 
\begin{equation} \label{smoothinglength}
h_{i}=\eta \left(\frac{m_{i}}{\rho_{i}}\right)^{1/3}, \ \ \ \ \ 
\end{equation}
where \textit{$\eta$} is a dimensionless parameter which determines the size of the smoothing length of the SPH particle given its mass and density. This relation is derived by requiring that a fixed mass, or equivalently a fixed number of neighbours, must be contained inside the smoothing sphere of each particle:
\begin{equation} \label{fixedmass}
\frac{4\pi}{3}\left(2 h_{i}\right)^{3} \rho_{i}=m_{i} N_{opt}=constant. \ \ \ 
\end{equation}
Here N$_{opt}$ denotes the number of neighbours inside the smoothing sphere, which we set equal to $50$ for the 3D simulations reported in this paper. Substituting Eq. \ref{smoothinglength} into Eq. \ref{fixedmass} allows us to determine $\eta$:
\begin{equation} \label{eta}
\eta=\left(\frac{3 N_{opt}}{32 \pi}\right)^{1/3}.
\end{equation}

Since Eq. \ref{density} and Eq. \ref{smoothinglength} depend on each other, we solve the two equations iteratively at each time step and for each particle. The evolution of the system of particles is computed using the second-order predict-evaluate-correct (PEC) scheme as implemented in GRADSPH, which integrates the SPH form of the equations of hydrodynamics with individual time steps for each particle. For more details and a derivation of the system of SPH equations we refer to \citet{b122}.

After the predictor-corrector step, the astrochemistry package KROME is invoked to solve the chemistry. The package includes an extensive chemical network to model the chemistry and cooling of primordial gas, which we will describe in detail in section 3. After initializing the species abundances as mass fractions of the various species, the mean molecular weight and the adiabatic index are calculated self-consistently on the basis of the chemical composition of the gas. The conversion of temperature to energy and vice-versa is computed on the basis of the equation of state used in GRADSPH-KROME.

We consider a spherical cloud which is modeled as a distribution of $1150709$ SPH particles. Our primordial cloud model has a total mass $M = 1.3041 \times 10^{4}$~M$_{\odot}$ and a radius $R = 2.169$~pc, with a corresponding initial density of $\rho_{i} = 8.650 \times 10^{-20}$~g cm$^{-3}$. The total mass inside the sphere is three times the Jeans mass of the system. The gas is initially isothermal with a temperature of $T = 300$~K. The gas is under solid body rotation. We explore values of the rotational parameter $\beta{}$ (the ratio of the rotational energy to the gravitational potential energy of the cloud) of $0\%$, $5\%$, $10\%$. The gas is also turbulent with the turbulent Mach number $\mathcal{M}$ set to one of the values $0.5$, $1.0$, $2.0$ to allow subsonic, transonic, and supersonic turbulence to be included in the models summarized in tables 1 \& 2. 

We inject a spectrum of incompressible turbulence into the initial conditions by adding the velocity of 1000 shear waves with random propagating directions to the initial velocity of each particle. The wavelength $\lambda$ of the shear waves is distributed uniformly between $0.001 R$ and $R$, while the amplitude $A$ of the waves follows a spectrum with $A \sim \lambda^{p}$, with the index $p$ taking a value of 0.5 in our models (M1 - M6) except model M7 where for the supersonically turbulent gas the index $p$ takes a value of 1.0. The amplitude of the resulting turbulent velocity field is then rescaled so that the RMS Mach number of the turbulent flow with respect to the initial isothermal sound speed equals the value in each model. Note that this procedure corresponds to decaying incompressible turbulence. We do not attempt to model driven turbulence in our models.

For the initial conditions described above, the dynamical time $R/c_s$ of the system is $4.17 \times 10^{5}$~yrs and is an estimate of the collapse time of the system. A second relevant timescale is the initial freefall time which is expressed as

\begin{equation} \label{freefalltime}
t_{ff}=\sqrt{\frac{R^{3}}{G M}},
\end{equation}

\noindent and becomes $t_{ff}=417.365$ kyrs for our models. 

The protostars which appear in our simulations are represented by sink particles which are allowed to form when the local density of the gas reaches the sink density threshold, which is set to $\rho_{\rm sink} = 1.0 \times 10^{-11}$~g cm$^{-3}$. The mass of sink particles which represent Pop.~III protostars keeps growing because of accretion of material from the parent gas cloud. In our set of reference models we follow the process of protostellar accretion until the total mass accreted by all protostars in a simulation reaches $200$~M$_{\odot}$. This procedure enables us to compare the outcome of the simulations at a similar evolutionary stage. The impact of varying the total mass contained within the sink particles as a criterion for terminating the simulations is discussed in section 4.8. The maximum density attained by the collapsing primordial gas in our simulations is $1.0\times10^{-11}$~g~cm$^{-3}$ which is around an order of magnitude larger when compared with the \cite{b24} simulation because our value for the threshold density for sink formation is one order of magnitude larger.

\begin{table*} \label{tbl-1}
\small 

\begin{flushleft}
\centering
\caption{Summary of the initial physical parameters and the final outcome of the simulation models. For each model, the initial radius, mass, average density and temperature are given by the constant values $2.169$~pc, $1.3041 \times 10^{4}$~M$_{\odot}$, $8.650 \times 10^{-20}$~g cm$^{-3}$ and $300$~K, respectively.}
\begin{tabular}{cccccc}
\hline
\hline
Model & Mach \# & Rotational factor ($\beta{}$ \%) & \# of objects & \# of binaries \\
\hline
M1  & 1.0   &  0  & 1  &  0\\
M2  & 0.5   &  0  & 1  &  0\\
M3  & 0.5	&  5  & 81 & 20\\
M4  & 1.0   &  5  & 38 & 10\\
M5  & 1.0   &  10 & 32 &  4\\
M6  & 0.5   &  10 & 36 &  6\\
M7	& 2.0   &  0  & 11 &  3\\

\hline
\end{tabular}
\end{flushleft}
\end{table*}

\begin{table*} \label{tbl-2}
\centering
 \caption{Summary of the maximum evolution time in units of the initial free fall time $t_{ff}$, the mass contained within proto-binary systems $M_{\rm binary}$, the mass contained within isolated protostars $M_{\rm isolated}$, the mean accretion rate of the most actively accreting proto-binary component $\dot M_{\rm acc.,most}(M_{\odot}$ yr$^{-1}$), and the mean accretion rate observed in the least active proto-binary component $\dot M_{\rm acc.,least}(M_{\odot}$ yr$^{-1}$) for models M1 - M7.}
 \begin{tabular}{cccccccc}
\hline
\hline
 Model          & M1    &    M2 &    M3 &    M4 &    M5 &    M6 &  M7 \\ 
\hline 
 $t_{ff}$       & 2.078924 & 2.073633 & 2.242504   & 2.24603   & 2.42592 & 2.42592 & 2.11512  \\
 $M_{binary}$   & - - - & - - - & 132.937 & 153.934 & 79.656 & 127.193 & 199.61  \\
 $M_{isolated}$ & 200 & 200 & 67.063 & 46.066 & 120.344 & 72.807 & 0.39  \\
 $\dot M_{\rm acc.,most}(M_{\odot}$ yr$^{-1}$) & 6.12 x 10$^{-1}$ & 6.57 x 10$^{-1}$  & 4.82 x 10$^{-2}$ & 1.95 x 10$^{-1}$ & 4.40 x 10$^{-2}$ & 7.25 x 10$^{-2}$ & 1.89 x 10$^{-1}$ \\ 
 $\dot M_{\rm acc.,least}(M_{\odot}$ yr$^{-1}$) & - - -              & - - -               & 2.44 x 10$^{-2}$ & 5.15 x 10$^{-2}$ & 1.75 x 10$^{-2}$ & 4.27 x 10$^{-2}$ & 8.80 x 10$^{-2}$ \\
\hline
\end{tabular}
\end{table*}

\subsection{Sink particle treatment and binary finder algorithm}

Our treatment of sink particles takes advantage of the recently improved sink particle algorithm for SPH calculations described by \citet{b126}. 
We introduce sinks into our current version of GRADSPH-KROME using the algorithm NEWSINKS by \citet[][see sec. 2]{b126}. 
In this algorithm, a sink particle can be created from a given SPH particle with accretion radius $r_{\rm acc}$ when 4 conditions are met: 

\begin{itemize}
\item its density exceeds the given threshold value $\rho_{sink}$. 
\item the gravitational potential of the particle is smaller than that of all of its neighbours, so that the particle sits in a gravitational potential minimum.
\item there is no overlap between the accretion radii of the potential new sink and any preexisting sinks.
\item The density of the particle satisfies the Hill condition 

\begin{equation} \label{Hillcondition}
\rho_{i} > \rho_{Hill}=\frac{3 X_{Hill}(-\Delta \vec{r}_{is^{'}}\cdot \Delta \vec{a}_{is^{'}})}{4\pi G \vert \Delta \vec{r}_{is^{'}} \vert^{2}},
\end{equation}

where $\Delta \vec{r}_{is^{'}}=\vec{r}_{i}-\vec{r}_{s^{'}}$ and $\Delta \vec{a}_{is^{'}}=\vec{a}_{i}-\vec{a}_{s^{'}}$ are relative displacement and acceleration vectors for 
the preexisting sinks labeled $s^{'}$, respectively, and $X_{Hill}$ is a user-defined parameter which we set equal to 4. Eq. (\ref{Hillcondition}) needs to be satisfied for all preexisting sinks in the simulation. The condition ensures that when a potential new sink particle is located in a region which already contains a sink at its center but which is much larger than the accretion radius of that sink (for example a region undergoing Kelvin Helmholtz contraction), a new sink can only be formed in the outer regions of the contracting region when there is a local density concentration that dominates the local gravitational field surrounding the potential new sink. 

\end{itemize}

The rate of accretion onto a sink particle is determined as follows. We assume spherical accretion onto the sink particle and for each particle $j$ whose distance from the sink particle is smaller than the accretion radius $r_{acc}$, we define its local accretion rate 

\begin{equation} \label{accretionrate}
\dot M_{j}=-4\pi \vert \Delta \vec{r}_{js} \vert \Delta \vec{r}_{js} \cdot \Delta \vec{v}_{js} \rho_{j},
\end{equation}

where $\Delta \vec{v}_{js}$ is the relative velocity vector between particle $j$ and sink $s$. 
Using the SPH smoothing kernel, we then define an accretion time scale $t_{acc}$ as

\begin{equation} \label{tacc}
t_{acc}=\frac{\sum_{j} m_{j} W_{j}}{\sum_{j} \vert \dot M_{j}\vert W_{j}}
\end{equation}

with $W_{j}=\frac{m_{j}}{\rho_{j}} W\left(\Delta \vec{r}_{js},H_{s} \right)$ and $H_{s}$ is the smoothing length of the sink which is set to half its accretion radius.
At each timestep $\Delta t$, the mass to be accreted by the sink is given by 

\begin{equation} \label{accretedmass}
\Delta M_{acc}=M_{int}\left[1-e^{-\Delta t/t_{acc}}\right],
\end{equation}

where $M_{int}$ is the total mass within the accretion radius of the sink. This accreted mass is removed from the surrounding SPH particles by looking first at the closest particle to the sink. If its mass it larger than the accreted mass, we subtract the accreted mass from its mass. Otherwise, we go to the second closest particle and continue until the accreted mass has been completely subtracted from the SPH particles within the interaction zone of the sink.
This algorithm is superseded in two cases. Firstly, we define a maximum mass $M_{max}$ as the mass of the SPH particles contained within the interaction zone at the moment the sink was created. If the mass within the interaction zone exceeds this value at any given time, we lower the accretion timescale by setting $t_{acc} \rightarrow t_{acc} \left(M_{max}/M_{int}\right)^{2}$. In this way a pileup of mass within the accretion radius is avoided and accretion is forced to be continuous. Secondly, if the timestep of a particle in the interaction zone satisfies $\Delta t_{j} < \gamma_{s} \sqrt{\frac{r_{sink}^{3}}{G M_{sink}}}$, the particle is accreted in its entirety without further checks. The parameter $\gamma_{s}$ is set to 0.01. 
The algorithm described above is called "regulated accretion". Further details and tests can be found in \citet{b126}.
The regulation of the accretion onto a sink particle described above is important because we do not apply boundary conditions at the sink accretion radius.

Because the mechanism for angular momentum transfer in accretion disks is still uncertain, we decided not to implement non-spherical disk accretion and angular momentum feedback from the sink particles (see sections 2.3.3 and 2.5 in \citet{b126}). 
We also do not consider mergers of sink particles which were taken into account by \citet{b43}.

In the plots shown in section 4 below, we compute and report the accretion rate onto sink particles by numerically differentiating the sink mass as a function of time. When referring to mean accretion rates, we consider time averages of the accretion rate computed in this way. We also note that the formation of sink particles leads to a decoupling between the sinks and the surrounding gas, as sink particles suddenly become insensitive to the hydrodynamic forces exerted by the surrounding gas. Since the hydrodynamical forces on a forming protostar which is strongly gravitationally bound are generally small, however, we do not think that this effect significantly affects the dynamics of the sink particles. As a result of this, the sinks are no longer coupled to the gas, and will in particular decouple from the center of the gas mass. Protostars which develop strong stellar winds should also push their surrounding gas aside, which would lead to even less pressure on the protostars (see for example \citealt{Dopcke13, b17, Kratter09}). Overall, it remains to be further explored what the best approximation would be to represent protostars into dynamical simulations, so that the coupling is neither too weak nor too strong. 

In order to determine pairs of sinks which form gravitationally bound binary systems, we define the total orbital energy per unit mass of a pair of sink particles as in \cite{b43}:

\begin{equation} \label{orbitalenergy}
\epsilon=\epsilon_{p}+\epsilon_{k},
\end{equation}

where $\epsilon_{p}$ and $\epsilon_{k}$ are the gravitational potential energy and kinetic energy per unit mass, respectively, and are defined as follows

\begin{equation} \label{potentialenergy}
\epsilon_{p}= -\frac{G\left(M_{1}+M_{2}\right)}{r},
\end{equation}

and

\begin{equation} \label{kineticenergy}
\epsilon_{k}=\frac{1}{2}v^{2},
\end{equation}

knowing that $r$ is their mutual distance, $v$ is their relative velocity, and $M_{1}$ and $M_{2}$ are the masses of the pair, respectively. A pair of sinks is considered a binary if $\epsilon < 0$.
Given the masses, the positions and the velocities of sink particles which are components of a binary, we determine the instantaneous orbital elements of the binary using standard formulae from celestial mechanics. We refer the reader to chapter 6 of \citet{Danby} for details.

\subsection{Resolution criteria}

Both the spatial and mass resolution play a vital role in self-gravitating SPH simulations using sink particles. In particular, we need to ensure two main criteria: The accretion radius of a fragment must be greater than the Jeans radius, and the mass of the fragment must be greater than the mass resolution adopted in the SPH scheme. These conditions are imposed to ensure that no unphysical fragmentation occurs and to prevent unphysical behaviour in SPH simulations of star formation \citep{b35}. 

We define suitable resolution criteria as follows. At a given density $\rho$, fragmentation of a local clump of gas can be prevented by a sufficiently strong gas pressure. However, gravitational collapse of the region will occur when the region has a characteristic radius greater than the Jeans length, 

\begin{equation} \label{Jeans Radius}
R_{J}=0.76 \left(\frac{RT}{G\rho}\right) ^{1/2},
\end{equation}

or equivalently, a mass greater than the Jeans mass, 

\begin{equation} \label{Jeans Mass}
M_{J}=1.86 \left(\frac{RT}{G}\right) ^{3/2} {\rho}^{-1/2},
\end{equation}

where $R$, $T$, $\rho$ and $G$ are the gas constant, the local temperature, the local density and the gravitational constant, respectively. The expressions for the Jeans radius and the Jeans mass are in cgs units.

When the density of an SPH particle exceeds the density threshold for the formation of sink particles and the additional conditions for sink particle creation described in section 3.1.1 are fullfilled, 
Eq.~(\ref{Jeans Radius}) is used to determine the accretion radius r$_{\rm acc}$ of the new sink particle. 
We set $r_{\rm acc} = 26.0$~au in our simulations, which is slightly above the Jeans radius at the moment of sink particle formation. 

In order to properly resolve fragmentation according to the mass resolution condition for SPH simulations defined by \citet{BateBurkert}, the Jeans mass evaluated at the position of each SPH particle must always satisfy the condition 
\begin{equation} \label{massresolutioncriterion}
M_{Jeans} > M_{\rm resolution}.
\end{equation}
$M_{\rm resolution}$ is defined as
\begin{equation} \label{Mass Resolution}
M_{\rm resolution}=2 N_{\rm opt} m_{\rm particle},
\end{equation}
with $N_{\rm opt}$ the number of neighbouring particles and m$_{\rm particle}$ the mass of a single particle in the SPH code. From Eq.~(\ref{Mass Resolution}) we calculate the mass resolution of our simulations by setting N$_{\rm opt} = 50$ and $m_{\rm particle}=0.0133 ~M_{\odot}$, where $m_{\rm particle} =1.3041 \times 10^{4} ~M_{\odot}/N_{\rm SPH}$ and $N_{\rm SPH} = 1150709$. For our simulations, we get $M_{\rm resolution} = 1.133 ~M_{\odot}$. We check that Eq. ~(\ref{massresolutioncriterion}) is never violated in our simulations.

\subsection{Chemistry and cooling}

The KROME package has already been employed to study a variety of astrophysical problems that cover a wide range of chemical and physical conditions \citep{Bovino2013, Katz2015, Prieto2015, Bovino2016, Schleicher2016, SeifriedandWalch2016, Capelo2017, Lupi2017, Kortgen2017}. In the present work, we employ a primordial network with the main cooling/heating processes required to model collapsing primordial gas. Our chemistry model includes a total of 9 chemical species: H, H$^{+}$, He, He$^{+}$, He$^{++}$, e$^{-}$, H$_{2}$, H$_{2}^{+}$, H$^{-}$. The initial abundances of these chemical species, expressed as mass fractions, are as follows: $f_\mathrm{H}=0.75$, $f_\mathrm{He} = 0.24899$, $f_\mathrm{H^+} = 8.2\times10^{-7}$,  $f_\mathrm{e} = 4.4 \times 10^{-10}$, and $f_\mathrm{H_2} = 10^{-3}$. All the other species are set to zero. We evolve the abundances of the chemical species contained within the gas by solving the rate equations using the DLSODES solver as outlined in \citet{b20}. We do not consider the radiative feedback from the protostars and neglect the possible presence of a UV background so that our simulations can be compared to \cite{b43}. The chemical reactions going on within the gas are described using the primordial network described in \citet{b20} in the absence of a radiative background. We use the H$_{2}$ cooling function provided by \citet{b34} and updated to \citet{Glover2015}. We also include continuum and Compton cooling as described in \citet{Omukai2001}, and consider the formation/destruction of H$_{2}$ and the associated energy sources/sinks that produce heating/cooling of the gas. 
\
	
\begin{figure*}
    \centering
    \includegraphics[angle=0,scale=0.275]{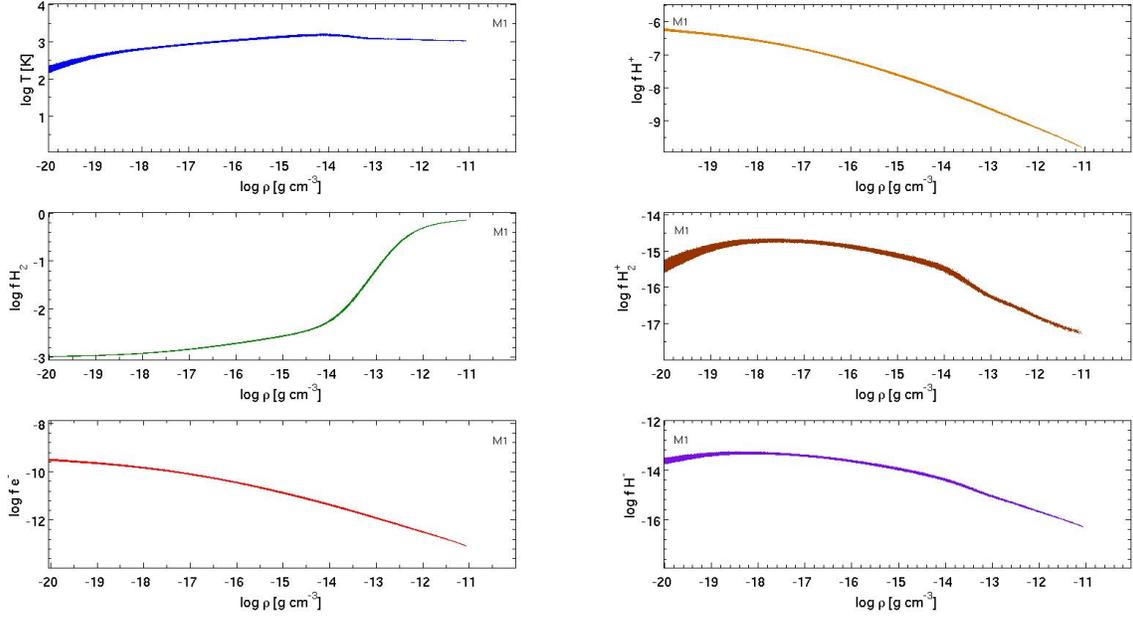}
    \caption{Evolution of thermodynamic quantities and chemical species as a function of log density for model M1 just before the formation of the sink particle. At the left, the temperature (K), the mass fraction of molecular hydrogen H$_{2}$, and the electron fraction e$^{-}$ are represented from top to bottom, respectively. At the right, the mass fractions of $H_{2}^{+}$, $H^{+}$, and $H^{-}$ are shown from top to bottom. Colour in online edition.}\label{fig:2}
  \end{figure*}

\begin{figure*} 
    \centering
    \includegraphics[angle=0,scale=0.275]{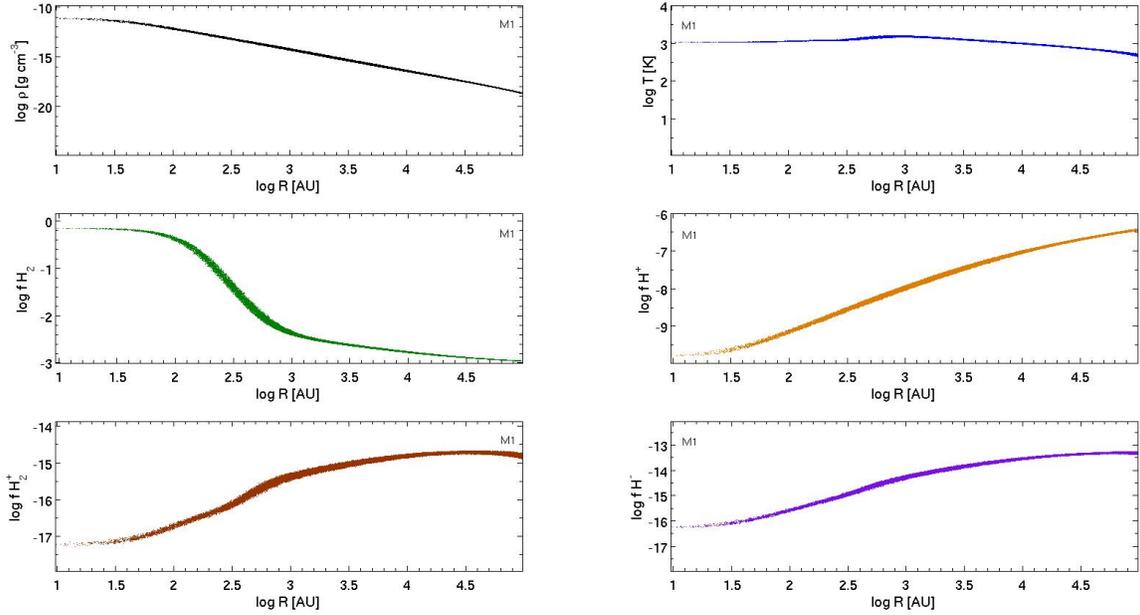}
    \caption{Radial profiles of the thermodynamic quantities and chemical species are shown for model M1 just before the formation of the sink particle. At the left, the density (g cm$^{-3}$) and the mass fractions of molecular hydrogen H$_{2}$ and $H_{2}^{+}$ are represented from top to bottom, respectively. At the right, the temperature and the mass fractions of $H^{+}$ and $H^{-}$ are shown from top to bottom. Colour in online edition.} \label{fig:3}
  \end{figure*}
  
  \begin{figure*}
\centering
\includegraphics[angle=0,scale=0.45]{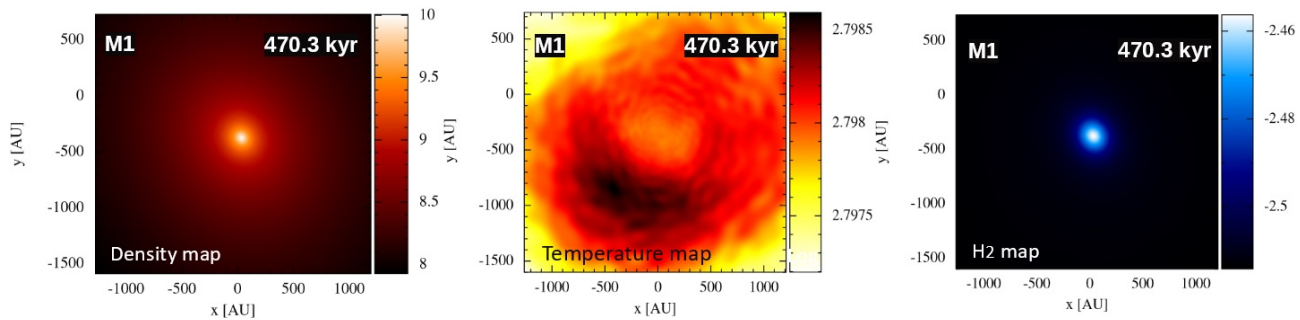}
\caption{Results of the simulation for model M1. The plots show face-on views of the gas column density, the temperature (slice plot), and molecular hydrogen column density integrated along the rotation axis at the times when the evolution of the models is terminated according to the stopping criterion. The respective colour bars on the right of each column show log ($\Sigma{}$) in g cm$^{-2}$, log T in K, and log (H$_{2}$). Each calculation was performed with 1150709 SPH particles. Colour in online edition.} \label{fig:1}
\end{figure*}
  

\section{Results and Discussion}

In this paper, we aim to investigate the effect of different turbulent Mach numbers $\mathcal{M}$ and rates of rotation $\omega{}$ on the process of fragmentation taking place in a Jeans unstable primordial gas cloud. $\omega{}$ is defined as the angular speed with which the gas sphere initially rotates as a solid body. 
The importance of the rotational energy is further quantified by the rotational parameter $\beta{}$ which is defined as the ratio of rotational energy to gravitational potential energy of the system. The evolution of our models leads to the formation of a rich population of Pop.~III protostellar systems. In the following sections, we describe and systematically discuss our results.

\subsection{Thermal and chemical evolution}
   
We first discuss the thermal and chemical evolution of our models. We start by focusing on the non-rotating and weakly turbulent gas model M1 and present the detailed chemical evolution of the collapsing primordial gas in this model.  We present a detailed study of the evolution of the chemistry to validate the results obtained from GRADSPH-KROME. We use the initial chemical abundances described in section~3, which are employed for all models M1 - M7. 

Figures \ref{fig:2} and \ref{fig:3} illustrate how the gas chemically evolves as a result of the subsonic regime of turbulence imposed. The gas is initially at $300$~K with a mass density of $8.650 \times 10^{-20}$~g cm$^{-3}$. These characteristic values of temperature and density are based on the micro-physics of H$_{2}$ cooling with H$_{2}$ acting as the dominant cooling agent and leave their imprint on the thermal evolution of the primordial gas \citep{b4}. As the gas collapses under self gravity the highest density region starts forming molecular hydrogen, which then provides cooling as can be seen in the panels of figure \ref{fig:1}. From left to right it shows the column density map, the temperature map, and the molecular hydrogen map. The top left panel in figure \ref{fig:2} illustrates how the temperature of the gas evolves as a function of the density. It can be seen that for densities of $10^{-20}$ - $10^{-14}$~g cm$^{-3}$, the temperature keeps increasing as the H$_{2}$ fraction is approximately constant, and the level populations are approximately thermalized. However, at just above $10^{-14}$~g cm$^{-3}$, the gas starts producing substantial amounts of H$_{2}$ due to three body reactions, and H$_{2}$ then provides sufficient cooling to slightly decrease the temperature. A sink particle representing a Pop.~III protostar forms at a density of $10^{-11}$~g cm$^{-3}$. 

The mid-left panel of figure \ref{fig:2} describes the evolution of the H$_{2}$ mass fraction as a function of increasing density of the collapsing gas. The onset of three-body molecular hydrogen formation can be seen at a density just above $10^{-14}$~g cm$^{-3}$. The sharp increase of the H$_{2}$ abundance, however, saturates when the density reaches 10$^{-12}$ g cm$^{-3}$ and hydrogen becomes fully molecular. The electron mass fraction is shown in the bottom-left panel of figure \ref{fig:2} and the mass fraction of H$^{+}$ is represented in the top-right panel of the same figure. The concentration of e$^{-}$ as well as H$^{+}$ steadily decreases with increasing density due to the effect of recombination. The evolving mass fractions of H$_{2}^{+}$ \& H$^{-}$ are presented in the mid-right and bottom-right panels of figure~1, respectively. In the initial phase of collapse (i.e. $10^{-20}$ - $10^{-14}$ g~cm$^{-3}$) these two chemical species show a rising trend up to a density of around $10^{-18}$~g cm$^{-3}$. Subsequently, their mass fraction falls down again, initially with a less steep slope followed by a much sharper decrease from a density of 10$^{-14}$ g~cm$^{-3}$ onwards. 

Figure \ref{fig:3} shows the radial profiles of the thermodynamic quantities of the gas as well as the chemical species involved at the time just before sink formation. The top-left and the top-right panels show how density and temperature evolve as a function of cloud radius. The density of the collapsing gas increases with decreasing radius and follows the expected profile of an isothermal collapse. The profile flattens around $5 \times 10^{-12}$~g cm$^{-3}$ within a radius of $25$~au from the central density peak. The temperature profile shown in the top-right panel shows a gradual increase of the temperature within a radius of $7000$~au from the point of maximum density. The temperature shows a small peak at around $1000$~au before three-body H$_{2}$ formation becomes efficient, subsequently decreases and finally reaches a plateau of around $1000$~K. Radial profiles of the mass fractions for H$_{2}$, H$_{2}^{+}$, H$^{+}$, and H$^{-}$ are presented in the mid-left, bottom-left, mid-right, and bottom-right panels, respectively. For H$_{2}$, initially a gradual increase in the mass fraction of H$_{2}$ is observed up to a radius of $1000$~au. From this point, three-body reactions start to produce a significant amount of H$_{2}$ and the mass fraction of H$_{2}$  increases until a radius of $50$~au from the central point of maximum density. Inside this region with a radius of $50$~au, the H$_2$ fraction saturates as the gas becomes fully molecular. 
The mass fraction of the two species H$_{2}^{+}$ \& H$^{-}$ first shows an increasing trend until a radius of $4 \times 10^{4}$~au, and a subsequently decreasing trend until a radius of $25$~au. These two species behave quite similarly in terms of their evolution of the mass fraction as a function of radius.  
On the other hand, the radial profile of the mass fraction of H$^{+}$ shows a different behaviour compared to that of H$_{2}^{+}$ and H$^{-}$. The H$^{+}$ fraction continuously decreases up to around $25$~au without a local maximum.

\subsection{Spherical collapse}
 
For gas clouds with an initially oblate shape, the angular momentum of the cloud almost never has an important impact on its collapse and fragmentation \citep{b36}. Our models instead consider a spherical distribution of gas within the primordial clouds, and the presence of angular momentum related to the rotation of the clouds produce different results than for an initially non-rotating gas cloud.  For a non-rotating cloud with $\beta{}$ = 0 and low levels of initial  turbulence $\mathcal{M}$ = 0.5 - 1.0, our primordial gas cloud models M1 \& M2 do not show any fragmentation and yield a single massive 200~$M_{\odot}$ Pop.~III star (see table 1). In each case, the Pop.~III protostar forms  at the very center of the collapsing cloud. This is mainly due to the absence of a disk structure because without rotational support the cloud collapses uniformly from all directions and gives birth to a single high density region which, after reaching the sink particle density, is converted into a single Pop.~III protostar. The protostar which appears in models M1 and M2 is formed at $470.74$~kyr and $469.54$~kyr, respectively. These individual protostars then start to accrete material continuously from the surrounding gas and gradually increase their masses. 

If a disk structure is formed in a rotating collapsing gas cloud and the protostar is part of the disk, accretion generally proceeds via disk instabilities \citep{b5}. However, in models M1 and M2 which are cases of initially non-rotating transonic and subsonic turbulent gas clouds, respectively, no such disk is formed during the collapse of the clouds despite the presence of net angular momentum related to the turbulence. The central protostar in models M1 \& M2 accretes a significant amount of gas in the radial direction. Model M1 needs about $1200$~yr longer than model M2 to create its first and only protostar (sink 1). However, it takes almost the same time of $2.0762$~$t_{ff}$ for the protostars in models M1 \& M2 to reach the mass limit of $200$~M$_{\odot}$ at which the simulations were terminated. It is thus interesting to note that models with Mach numbers up to $1$ do not produce strong deviations from a spherical collapse, as long as there is no net initial rotation. As shown below, such deviations occur however for a Mach number of $2$ in model M7 which involves a higher amount of turbulent angular momentum compared to models M1 \& M2.

\subsection{Rotational effect on fragmentation}

\begin{figure*} 
    \centering
    \includegraphics[angle=0,scale=0.775]{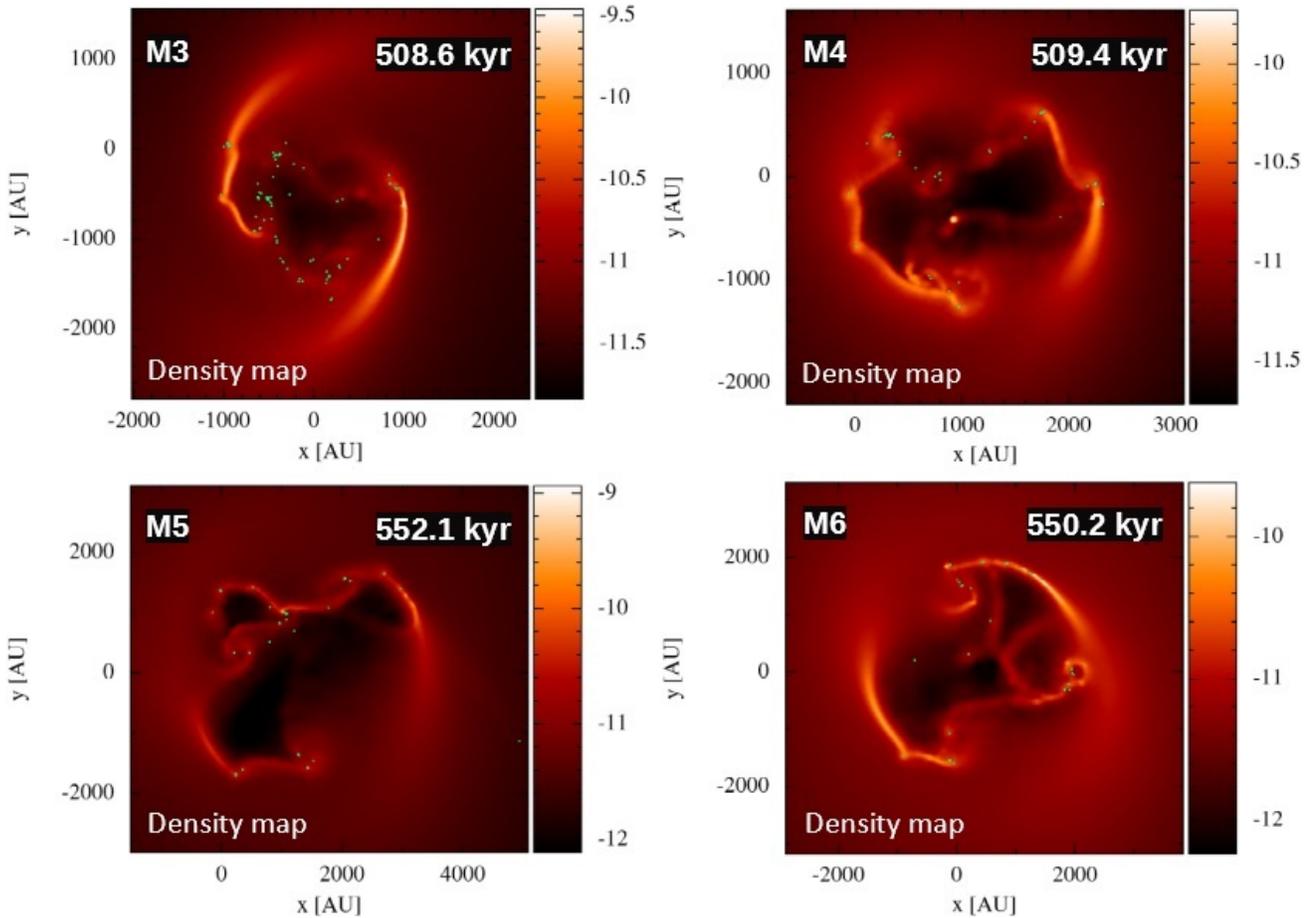}
    \caption{Results of the simulation for models M3, M4, M5, and M6. The plots show face-on views of the column density for the total gas density integrated along the rotation axis(z) at the time when the evolution of the models is terminated. The colour bars on the right of each panel show the logarithm of the column density ($\Sigma{}$) in physical units of g cm$^{-2}$. Each calculation was performed with 1150709 SPH particles. Colour in online edition.}\label{fig:4}
  \end{figure*}

\begin{figure*} 
    \centering
    \includegraphics[angle=0,scale=0.775]{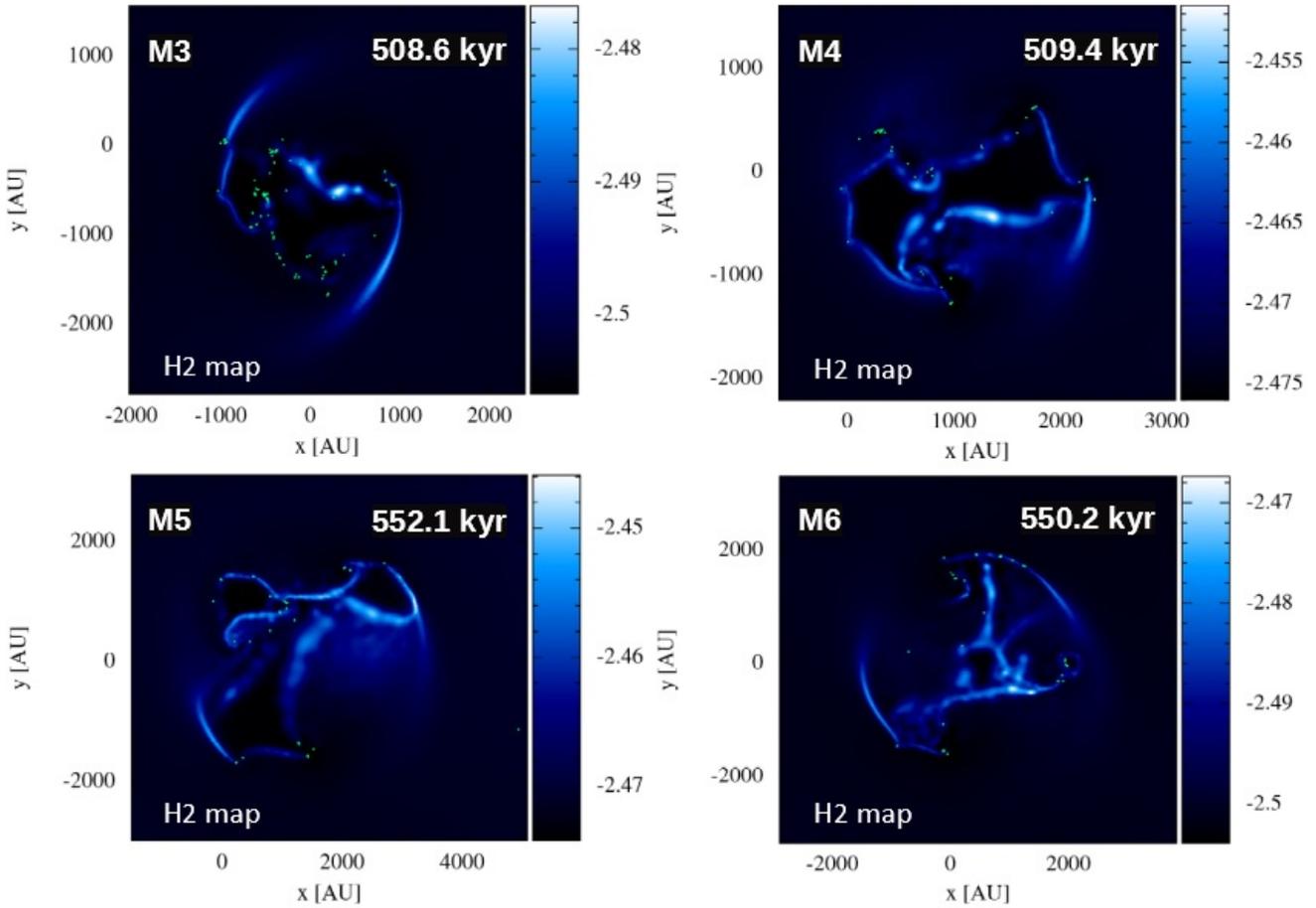}
    \caption{Results of the simulation for models M3, M4, M5, and M6. The plots show face-on views of the logarithm of the mass fraction of molecular hydrogen integrated along the rotation axis(z) at the time when the evolution of the models is terminated. The colour bars on the right of each panel show the logarithm of the mass fraction of H$_{2}$. Each calculation was performed with 1150709 SPH particles. Colour in online edition.}\label{fig:5}
  \end{figure*}

\begin{figure*} 
    \centering
    \includegraphics[angle=0,scale=0.55]{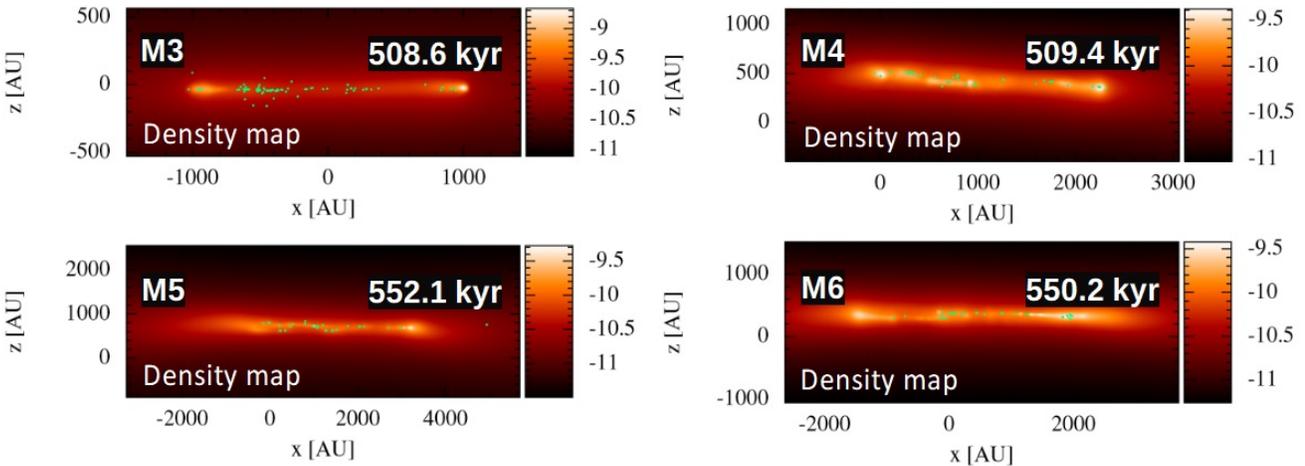}
    \caption{Results of the simulation for model M3, M4, M5, and M6. The plots show edge-on views of the column density in the x-z plane integrated along the y-axis at the time when the evolution of the models is terminated. The colour bars on the right of each panel show the logarithm of the column density ($\Sigma{}$) in physical units of g cm$^{-2}$. Each calculation was performed with 1150709 SPH particles. Colour in online edition.}\label{fig:6}
  \end{figure*}

The initial density at which our minihalo models initiate their collapse corresponds to a state where the collapse is already baryon dominated \citep{Loeb94, Oliveira98, b29, Susa2006, Schleicher09, Clark11}.  
Angular momentum also contributes to the process of fragmentation. Conservation of angular momentum leads to the formation of a rotationally flattened region where density perturbations can cause fragmentation with multiple centers of collapse \citep{b37,b38}. In our simulations, we find that a substantial amount of fragmentation occurs in the presence of initially non-zero rotation rates $\omega{}$ of the primordial gas cloud models. Figures \ref{fig:4} \& \ref{fig:5} show the evolved state of the four models M3, M4, M5 and M6 with initial rotation. For each model, we show a column density map for the total gas density (Figure \ref{fig:4}) and for molecular hydrogen (Figure \ref{fig:5}) in the x-y plane. In addition, we show in Figure \ref{fig:6} the corresponding column density maps in the x-z plane of these models to illustrate the disk structure of the collapsing gas. As summarized in table~1, the four primordial gas cloud models which are initially rotating with different rates also cover a range of turbulent Mach numbers from subsonic to the transonic turbulence, $\mathcal{M}=0.5, 1.0$. We find an interesting trend in the level of fragmentation when comparing models (M3 \& M6) and (M4, \& M5). Each set of models has identical initial turbulent Mach numbers $\mathcal{M} = 0.5$ and $\mathcal{M} = 1.0$, but have different ratios of rotational energy to gravitational potential energy $\beta{} = 5\% - 10\%$. The level of fragmentation seems to respond to the rate of rotation of the primordial gas clouds. We observe less fragmentation in the disk of model M6 which has $\beta{} = 10\%$ and therefore stronger rotational support. The first set of models  M3 \& M6 produces $81$ and $36$ protostars,respectively, while the second set of models M4 \& M5 yielded only $38$ and $32$ protostars which clearly reflects the difference in rotational energy between the two sets of models. 

It is interesting to compare the time taken by the two sets of models to reach the prescribed mass limit of $200$~M$_{\odot}$. In the first set of models M3 \& M6, the two clouds need $2.2425$~$t_{ff}$ and $2.4259$~$t_{ff}$, respectively, to reach the stage of evolution where the protostars collectively have a total mass of $200$~M$_{\odot}$. In the second set of models M4 \& M5 the two clouds with a respective total number of 38 and 32 protostars needed $2.2460$~$t_{ff}$ and $2.4259$~$t_{ff}$, respectively, to reach the mass limit of $200$~M$_{\odot}$. This observation confirms that Pop.~III protostars forming in clouds with stronger rotational support require more time to accrete significant amounts of gas from their surrounding envelope. This process does not seem to depend strongly on the number of protostars which are accreting gas from the envelope. Model M3 with $81$ protostars and model M4 with $36$ protostars take almost the same time to reach the mass limit of $200$~M$_{\odot}$ (see tables~1 and 2 for a summary).
Another interesting observation is that for the models with identical $\beta{}$ parameter and increasing turbulent Mach number $\mathcal{M}$, we see a decline in the total number of protostars formed during cloud collapse. In contrast, model M7 produces only 11 Pop.~III protostars and the limiting mass of $200$~M$_{\odot}$  is reached $0.2190$~$t_{ff}$ earlier compared to the rotating models.  

As can be seen in Figure~\ref{fig:6}, the disk structures appear slightly inclined with respect to the initial axis of rotation in models M4 and M5. This shift of the disk orientation is indeed observed for individual protostellar disks in models of collapsing transsonical turbulent prestellar cores \citep{b200}. On the other hand, no such tilt is observed in the disks formed in models M3 and M6. Despite the fact that models M4 and M5 cover both high and low values of $\beta{}$, the appearance of inclined disks can be explained by considering the relatively stronger initial turbulence with Mach number $\mathcal{M} = 1$ in this set of rotating models. The net contribution of the transonic turbulence leads to a random shift of the rotational axis away from the initial direction of rotation in models M4 and M5, and as a result the disks hosting the protostars appear with a slight tilt from the initial axis of rotation of the cloud.

\begin{figure*}
    \centering
    \includegraphics[angle=0,scale=0.55]{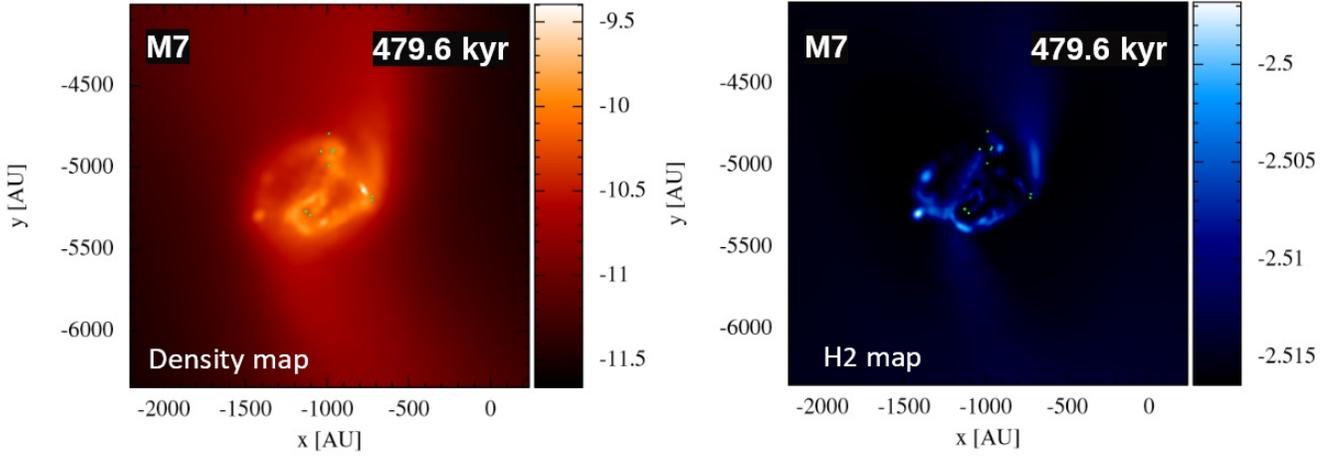}
    \caption{Results of the simulation for model M7. The plots show face-on views of the gas column density and the column density of molecular hydrogen integrated along the rotation axis(z) at the time the evolution of the models is terminated. The colour bars on the right of each panel show the logarithm of the column density ($\Sigma{}$) in physical units of g cm$^{-2}$ and the logarithm of the mass fraction of H$_{2}$. Each calculation was performed with 1150709 SPH particles. Colour in online edition.} \label{fig:7}
  \end{figure*}

Figure \ref{fig:7} shows the final outcome of model M7, which is a non-rotating case like models M1 \& M2, but with a supersonic level of initial turbulence ($\mathcal{M} = 2.0$). The plots in Figure \ref{fig:7} show a more strongly turbulent gas structure. As shown in table~1, the supersonic turbulence has produced a total of $11$ Pop.~III protostars. The total mass contained inside these $11$~protostars took around $2.11512$~$t_{ff}$ to approach $200$~M$_{\odot}$. As expected, regions with relatively cold gas collapse to higher densities and reach the sink formation density at various locations. The thick filamentary structure within the gas is the birth place for the Pop.~III protostars. Unlike the other two less turbulent models M1 \& M2, the supersonic gas in model M7 needs $0.0389$~$t_{ff}$ (162.355 kyr) longer to fragment to a level where the mass inside all protostars reaches $200$~M$_{\odot}$.

\subsection{Thermal response to $\mathcal{M}$ \& $\beta{}$}

\begin{figure*} 
    \centering
    \includegraphics[angle=0,scale=0.575]{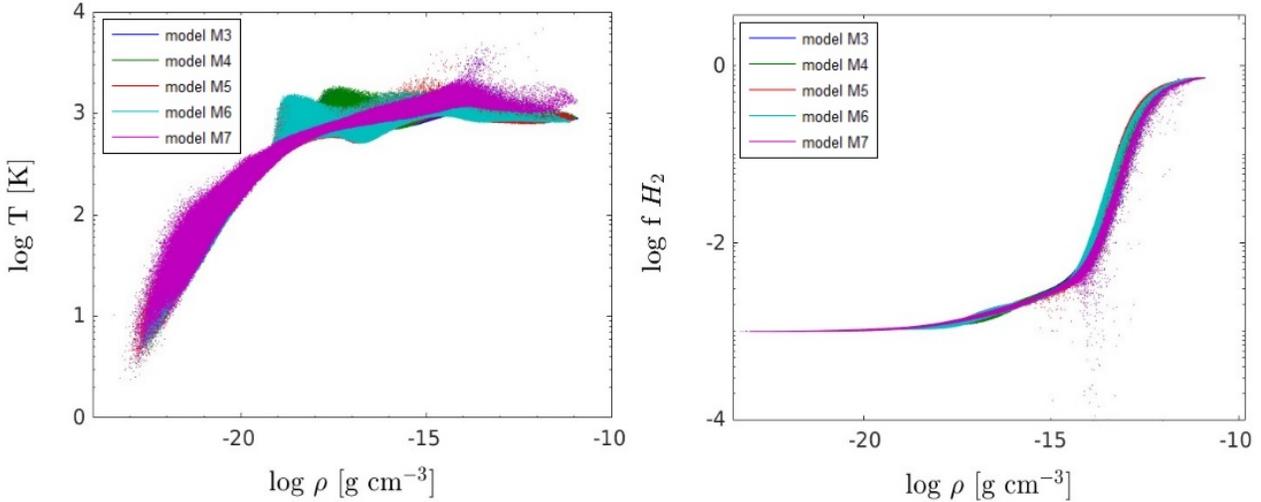}
    \caption{Phase diagrams of log T versus log $\rho{}$ (left panel) and the logarithm of the mass fraction of $H_{2}$ versus log $\rho{}$ (right panel) for models M3-M7 at the end of the simulations. Density is measured in g cm$^{-3}$ and temperature is measured in K. Colour in online edition. }\label{fig:8}
  \end{figure*}

Figure \ref{fig:8} shows phase diagrams for the temperature (log T versus log $\rho{}$) and the mass fraction of H$_{2}$ (log mass fraction of H$_{2}$ versus log $\rho{}$) for the rotating set of primordial gas cloud models at the time when the first sink particle is formed in these simulations. 
We present the diagrams for model M7 (a non-rotating case) along with the other models with initial rotation (M3 - M6) considering that model M7 has yielded multiple fragments including a few proto-binary systems in agreement with the rotating models. All of the models shown present more or less identical trends in both the thermal and molecular hydrogen diagrams. The left panel in Figure \ref{fig:8} illustrates the thermal behaviour for the five models. The collapsing gas mainly cools due to molecular hydrogen which starts forming via (3-body processes) \citep{b1,b39, Turk2010, Forrey2013, Bovino2k14} at a density of around $1.0 \times 10^{-14}$~g cm$^{-3}$ and a temperature below $2000$~K. At lower densities, H$_{2}$ may form via gas-phase reactions, which is however known to be inefficient, and therefore leads only to low abundances (see for example \citealt{b29, Omukai2005, Glover09}).
 It is important to note that in model M7 (the non-rotating case) the profile looks slightly smoother than the rest of the models (the rotating cases). 
  
\subsection{Evolution of clump masses}

At the time of their formation, Pop.~III protostars posses a mass less than 10$^{-2}$~M$_{\odot}$ \citep{b1,b6}. However, because of the large reservoir of surrounding gas and appropriate conditions for efficient mass accretion, these protostars can significantly increase their masses over time \citep{b41}. The sink particles in our simulations have a mass of $1.13 \times 10^{-2}$~M$_{\odot}$ at the time of their formation. They then begin to grow as the accretion continues. Figures~\ref{fig:9} \& \ref{fig:10} illustrate how the mass accretion of individual fragments proceeds. Figure \ref{fig:9} presents mass accumulation profiles for the protostars in models M1, M2 and M7. The top and bottom panels in the left part of the figure show the mass accretion histories for the single sink particle (sink 1) which is formed in models M1 and M2. In each model, the sinks have a fairly similar history of mass accretion from their parent gas cloud, apart from the difference of around $\approx{} 1200$~yrs in the time required for their formation. The sinks in each model attain $200$~M$_{\odot}$ by the time the simulations are terminated at $470.30$~kyr and $470.03$~kyr, respectively. The top \& bottom panels in the right part,on the other hand, are for the sink particles labeled sink~1 and sink~5 in model M7, which are the most actively and the least actively accreting components of proto-binary systems in this model (see table 7 for reference). By the time terminate the simulation is terminated, sink 1 and sink 5 in model M7 have masses of $67.50$~M$_{\odot}$ and $12.06$~M$_{\odot}$.

Figure \ref{fig:10} gives an overview of the mass accretion history for selected sink particles in the models M3 - M6 which have an initial rigid-body rotation. We have selected the sink particles which are members of proto-binary systems with the largest and smallest accretion rates, respectively. The blue and green curves in the top and the bottom parts of Figure \ref{fig:10} show the mass accretion histories of the most actively accreting and the least actively accreting members of proto-binary systems in the four indicated models, respectively. The most actively accreting sink particle in model M3 ($\mathcal{M}$ = 0.5 \& $\beta{}$ = 5\%) is sink 41. Its mass accretion history is shown in the top left panel of the upper part in Figure \ref{fig:10}. After its formation at a time of $507.4$~kyr, sink 41  gradually increases its mass. However, after an initially rather gradual growth, a sudden bump occurs around $507.80$~kyr after which the mass increases more quickly. A final mass of $1.45$~M$_{\odot}$ is reached at the end of the simulation. 

The bottom left panel of the upper part of the same figure shows the mass accumulation history of sink 3, the most actively accreting binary component in model M4 ($\mathcal{M}$ = 1.0 \& $\beta{}$ = 5\%) which is formed at a time of $507.75$~kyr. This system shows no sudden increase in its accretion history, but keeps acquiring mass from the surrounding gas in a more gradual manner. It has gained a mass of $14.26$~M$_{\odot}$ when the simulation is finished. The top right panel shows the mass accretion history of the most actively accreting binary component sink 13 in model M5 ($\mathcal{M} = 1.0$ \& $\beta{} = 10\%$). After the formation of sink 13 at $549.8$~kyr, it shows a gradually increasing mass leading to a final mass of 6.0 $M_{\odot}$ when the simulation ends. Finally, the bottom right panel shows the accretion history of sink 6 in model M6 ($\mathcal{M}$ = 0.5 \& $\beta{} = 10\%$), which is selected according to the same criteria. We observe a more prominent rising trend in the mass accumulation history of sink 6. From its formation time at $547.90$~kyr it keeps accreting material from the surrounding and attains a final mass of $1.34$~M$_{\odot}$ at the end of the simulation.

In a similar way, the top left panel of the bottom part of Figure \ref{fig:10} ilustrates the mass accumulation history for sink 34, the least actively accreting sink among all of the proto-binary systems in model M3 ($\mathcal{M} = 0.5$ \& $\beta{} = 5\%$). It is formed at $507.30$~kyr and acquires a final mass of $2.62$~M$_{\odot}$. 
The least actively accreting proto-binary members in models M4, M5 and M6 are sinks 4, 8 and 24, respectively.
Sink 4 is formed at $507.8$~kyr and grows to a final mass of 17.74 $M_{\odot}$.  Sink 8 is formed  at $549.70$~kyr and acquires a final mass of $26.34$~M$_{\odot}$. Sink 24 is created at $548.50$~kyr and reaches a mass of  $4.36$~M$_{\odot}$ at the end of the simulation.

\subsection{Accretion rates of Pop.~III protostars}

\begin{figure*} 
    \centering
    \includegraphics[angle=0,scale=0.35]{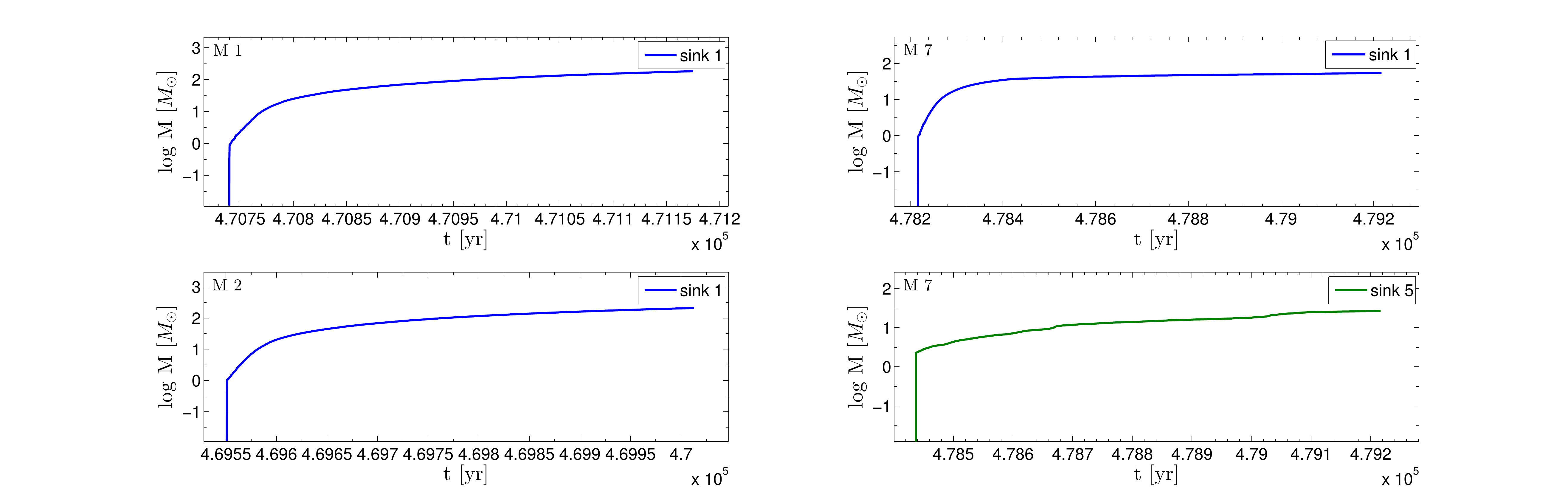}

    \caption{Mass accumulation histories for the most actively accreting binary component (top right panel) and the least actively accreting binary component (bottom right panel) in model M7. The top and bottom left panels show the mass accretion history for the sink particles in models M1 \& M2, respectively. The accumulated mass is represented in solar mass units $M_{\odot}$ and the time is expressed in years. Colour in online edition.} \label{fig:9}
     \end{figure*}

The mass accretion rate and its time evolution play a sensitive role in determining the mass of Pop.~III stars forming in primordial gas \citep{b31}. The process of  accretion also continues during the entire lifetime of these protostars \citep{b31, b44, b173, b45}. In Figures \ref{fig:11} \& \ref{fig:12} we present a general overview of the mass accretion rates for the protostars formed in models M1 - M7. In each panel, the average accretion rate is indicated and summarized in table 2.  
The values for the average accretion rate are calculated including the initial peak in the profiles for the accretion rate. As time proceeds, a general decreasing trend in the mass accretion rates is observed. Figure \ref{fig:11} illustrates the non-rotating models M1, M2, and M7 is complementary to Figure \ref{fig:9}. Because the angular momentum of the gas plays an important role it is expected that mass accretion only continues if rotation does not halt the collapse \citep{b42}. 

The highest accretion rates are clearly observed in the non-rotating models. The top left panel shows the mass accretion rate associated with the only sink particle which appears in model M1. The mass accretion rate for this protostar which has formed in transonically turbulent gas keeps fluctuating around a mean value of $6.12 \times 10^{-1}$~M$_{\odot}$yr$^{-1}$ (see table 2). The bottom left panel of the same figure represents the mass accretion rate for the sink particle in model M2. Here the turbulence is subsonic and the protostar accretes material from its surroundings at a mean rate of $6.57 \times 10^{-1}$~M$_{\odot}$yr$^{-1}$. The top right and the bottom right panels respectively represent the mass accretion rates of the most actively accreting and the least actively accreting Pop.~III proto-binary components in model 7. Model 7 is initialized with supersonic turbulence and the most actively accreting protostar (sink 1) in the top right panel exhibits a high accretion rate which peaks at about $7.03 \times 10^{-1}$~M$_{\odot}$yr$^{-1}$  early in the course of the evolution of this model. However, as the simulation proceeds until $479.20$~kyr, we observe a considerable decline of the mass accretion rate to $2.210 \times 10^{-2}$~M$_{\odot}$yr$^{-1}$. Nevertheless, the mean accretion rate keeps oscillating around a value of $1.89 \times 10^{-1}$~M$_{\odot}$yr$^{-1}$ over the entire lifetime of the sink particle. The bottom right panel of the same figure shows the history of the accretion rate of the least actively accreting proto-binary component (sink 5) in model M7. This protostar accretes material from its surroundings with a mean accretion rate $8.80 \times 10^{-2}$~M$_\odot$yr$^{-1}$. Being the least active binary component in terms of its mass accretion, the mass accretion rate for this Pop.~III protostar drops down to a minimum value of $2.05 \times 10^{-2}$~M$_{\odot}$yr$^{-1}$ as the system evolves up to $479.20$~kyr.

\begin{figure*} 
\centering
     \includegraphics[angle=0,scale=0.35]{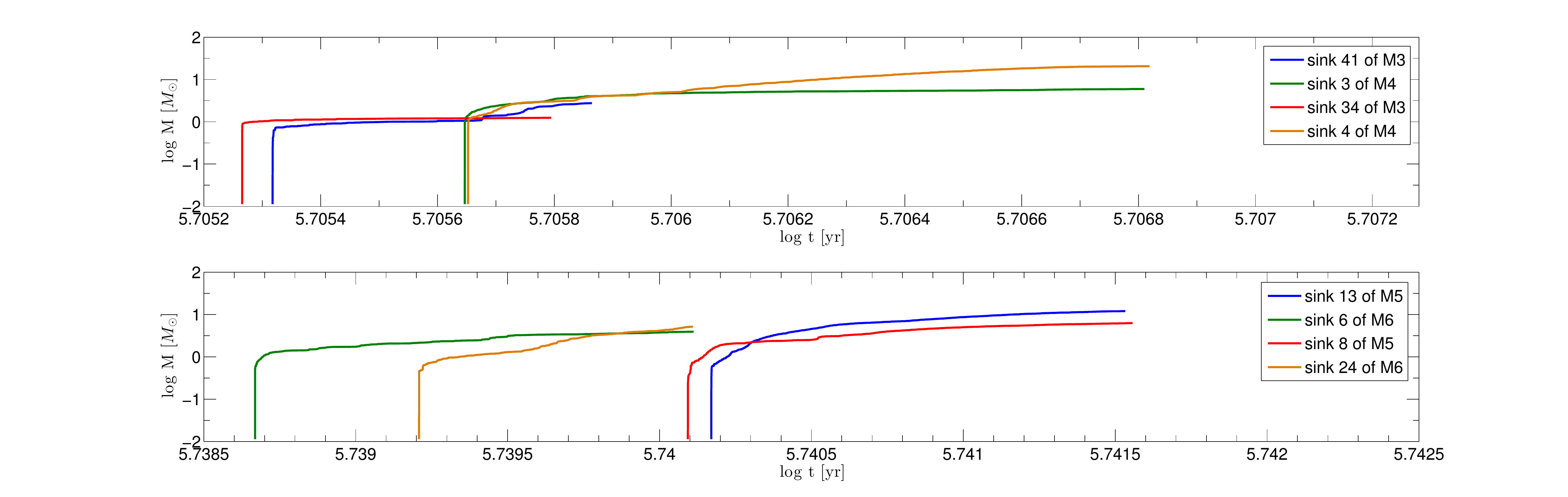} 
\caption{Top - mass accumulation profiles of the most actively accreting proto-binary components (blue and green curves) and the least actively accreting proto-binary components (red and orange curves) for models M3 and M4. Bottom - mass accumulation profiles of the most actively accreting proto-binary components (blue and green curves) and the least actively accreting proto-binary components (red and orange curves) for models M5 and M6. The mass is represented in solar mass units ($M_{\odot}$ ) and the time is expressed in years. Colour in online edition.}\label{fig:10}
\end{figure*}

\begin{figure*} 
\centering
\includegraphics[angle=0,scale=0.35]{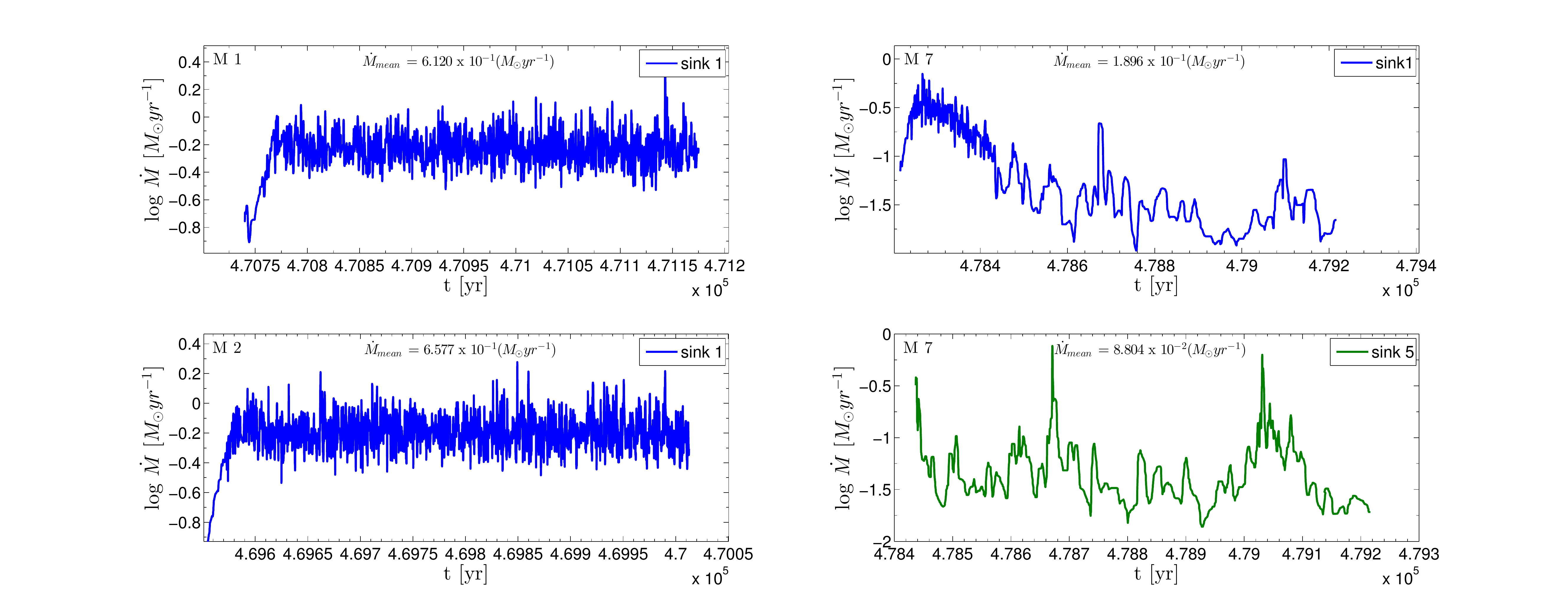}   
\caption{Evolution of the mass accretion rate for the most actively accreting proto-binary component (top right panel) and the least actively accreting proto-binary component (bottom right panel) in model M7. The top and bottom left panels show the evolution of the mass accretion rate for the sink particles in models M1 \& M2, respectively. The accretion rate is given in units of solar mass per year and the time is mentioned in years. Colour in online edition.}\label{fig:11}
\end{figure*}

\begin{figure*} 
    \centering
       \includegraphics[angle=0,scale=0.35]{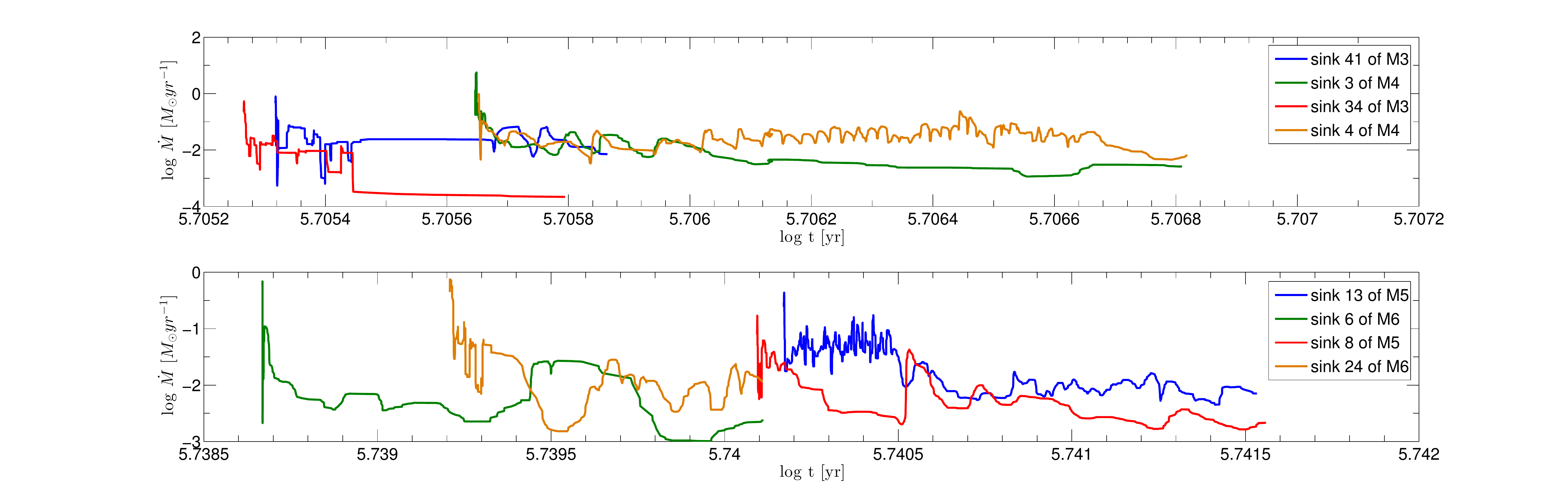}
  \caption{ Top - mass accretion rate profiles of the most actively accreting proto-binary components (blue and green curves) and the least actively accreting proto-binary components (red and orange curves) for models M3 and M4. Bottom - mass accretion rate profiles of the most actively accreting proto-binary components (blue and green curves) and the least actively accreting proto-binary components (red and orange curves) for models M5 and M6. Accretion rates are given in units of solar mass per year and the time is mentioned in years. Colour in online edition.}\label{fig:12}
     \end{figure*}

Figure \ref{fig:12} illustrates the mass accretion history for the models M3 - M6 which have initial rotation. We again follow the same criteria to select the protostars as discussed previously in relation to Figure \ref{fig:11} and Figures \ref{fig:9} and \ref{fig:10}. The colour scheme is also identical to that used previously in Figure \ref{fig:10}. Focusing on the curves in the top part, the top left panel of Figure \ref{fig:12} shows the results for sink 41 in model M3 ($\mathcal{M} = 0.5$ \& $\beta{} = 5\%$) which has a mean accretion rate of $4.82 \times 10^{-2}$~M$_{\odot}$yr$^{-1}$. 

However, the later stages of its evolution which goes until $508.0$~kyr show that even for this most active protostar, the accretion rate drops toward $7.12 \times 10^{-3}$~M$_{\odot}$yr$^{-1}$. The bottom left panel shows the mass accretion history of sink~3 in model M4 ($\mathcal{M} = 1.0$ \& $\beta{} = 5\%$). This protostar has an average accretion rate of $1.953 \times 10^{-1}$~M$_{\odot}$yr$^{-1}$. As the system evolves up to $509.10$~kyr the accretion rate drops down to $2.59 \times 10^{-3}$~M$_{\odot}$yr$^{-1}$. 
The top right panel presents the results for sink 13 in model M5 ($\mathcal{M} = 1.0$ \& $\beta{} = 10\%$). For this protostar, frequent high accretion peaks are observed initially. But when the simulation proceeds to $551.50$~kyr, the accretion rate drops to $7.44 \times 10^{-3}$~M$_{\odot}$yr$^{-1}$. With a very active initial accretion phase, this protostar has a mean accretion rate of $4.40 \times 10^{-2}$~M$_{\odot}$yr$^{-1}$. 
The bottom right panel shows the mass accretion history for sink 6 in model M6 ($\mathcal{M} = 0.5$ \& $\beta{} = 10\%$). 
In this case the protostar keeps accreting material from the surrounding gas at an average rate of $7.25 \times 10^{-2}$~M$_{\odot}$yr$^{-1}$. At a later stage of its evolution when the simulation has advanced to $549.70$~kyr, however, the mass accretion rate drops to $2.37 \times 10^{-2}$~M$_{\odot}$yr$^{-1}$, but the accretion rate again shows a rising trend until the simulation is finally terminated. 

On the bottom part of Figure \ref{fig:12}, we show for each model the time evolution of the accretion rates of the Pop.~III proto-binary components which are the least active in terms of their mass accretion rates. The top left panel shows sink 34 in model M3 ($\mathcal{M} = 0.5$ \& $\beta{} = 5\%$). 
Initially this protostar accretes at a high rate of $1.36 \times 10^{-2}$~M$_{\odot}$yr$^{-1}$ until $507.50$~kyr. Later on the accretion rate drops considerably to $2.18 \times 10^{-4}$~M$_{\odot}$yr$^{-1}$ until the simulation ends at $507.9$~kyr. The mean accretion rate for this protostar is $2.44 \times 10^{-2}$~M$_{\odot}$yr$^{-1}$. 
The bottom left panel illustrates the results for sink 4 in model M4 ($\mathcal{M} = 1.0$ \& $\beta{} = 5\%$) which has mean accretion rate of $5.15 \times 10^{-2}$~M$_{\odot}$yr$^{-1}$. This proto-binary component also accretes material at a relatively higher rate of $2.31 \times 10^{-1}$~M$_{\odot}$yr$^{-1}$ until $508.7$~kyr and subsequently shows a drop of its mass accretion rate to $6.43 \times 10^{-3}$~M$_{\odot}$yr$^{-1}$ at $509.1$~kyr. 
The top right panel shows the accretion history for sink 8 in model M5 ($\mathcal{M} = 1.0$ \& $\beta{} = 10\%$). This protostar generally exhibits a low level of accretion activity with an average accretion rate of $1.75 \times 10^{-2}$~M$_{\odot}$yr$^{-1}$, except for the earlier stage at $550.2$~kyr where the protostar shows a relatively high mass accretion rate of $4.25 \times 10^{-2}$~M$_{\odot}$yr$^{-1}$ which quickly declines. The mass accretion reaches its lowest value of $2.15 \times 10^{-3}$~M$_{\odot}$yr$^{-1}$ at $551.5$~kyr. Finally, the bottom right panel of Figure 12 presents the mass accretion history of sink 24 in model M6 ($\mathcal{M} = 0.5$ \& $\beta{} = 10\%$), which has a mean accretion rate of $4.27 \times 10^{-2}$~M$_{\odot}$yr$^{-1}$. The accretion history of this protostar also shows, in general, a decreasing trend, with a peak as high as $1.30 \times 10^{-1}$~M$_{\odot}$yr$^{-1}$ at $548.6$~kyr in the earlier phase of the evolution, after which a decline follows to $1.77 \times 10^{-3}$~M$_{\odot}$yr$^{-1}$ when the system evolves up $549.0$~kyr. Finally there is again an accretion phase which involves an increase in the mass accretion rate to $2.81 \times 10^{-2}$~M$_{\odot}$yr$^{-1}$ at $549.10$~kyr. The mass accretion rate then keeps on fluctuating until the simulation ends at $549.7$~kyr with a final mass accretion rate of $1.28 \times 10^{-2}$~M$_{\odot}$yr$^{-1}$.  

We can compare the accretion rates reported in this section to the previous work in the literature. For the simulations with initial angular momentum illustrated in figures 10 and 12, most of the mass growth of the sinks has happened within a timescale of 1000-2000 yrs, although the protostars are still gaining substantial mass at the end of the simulations. 
On average the range of the accretion rates at the end of the runs ($\sim 1.0 \times 10^{-4}$~M$_{\odot}$yr$^{-1}$ to $\sim 1.0 \times 10^{-2}$~M$_{\odot}$yr$^{-1}$) and the protostellar masses are quite similar to the previous work of \citet{b24} (their Figure 9) and \citet{b43} (their Figure 5). The effect of radiative feedback is illustrated in Figure 10 of \citet{Stacy2012} and implies an only moderate difference of 25 $\%$ in the mass of the first sink formed in their simulation. On the other hand, \citet{Susa2013} simulated the growth of protostars in a collapsing minihalo up to the main sequence phase over a timescale of $10^{5}$ yrs using an RSPH code. We expect that his runs without feedback reported in his Fig. 5 are the closest to the results of our paper. We generally estimate in agreement with \citet{Susa2013} that sink particle schemes will tend to overestimate the growth rate of protostars certainly if pressure effects at the accretion radius are not explicitly taken into account. The models M1, M2 and M7 with zero initial angular momentum have considerably higher accretion rates, but we do not think that zero angular momentum models, although of theoretical importance, are realistic approximations of the initial conditions for Pop III star formation.

\begin{table*} \label{tbl-3}
\centering
 \caption{Summary of the results for model M3. The columns include the indices of the sink particles which are components of binaries, the masses of the components, the semi-major axis (a), the eccentricity (e), the mass ratio (q), and the binary separation (d) at the end of the simulation.} 
 \begin{tabular} {ccccccc}
\hline
\hline
 Component indices & component masses ($M_{\odot}$) & a (au) & e &  q  &  d(au) \\
 \hline
sink 2, sink 3	&	12.83, 11.83		&	3.08		&	0.95		&	0.92		&	  6.02 \\
sink 3, sink 24	&	11.83, 1.07		&	43.06	&	0.48		&	0.009	&	 62.86 \\
sink 4, sink 31	&	3.35	,  1.40		&	77.81	&	0.82		&	0.42		&	116.50\\
sink 8, sink 5	&	13.34, 3.23		&	22.72	&	0.97		&	0.24		&	 34.54\\
sink 6, sink 34	&	17.15, 2.62		&	29.04	&	0.74		&	0.15		&	 50.36\\
sink 7, sink 31	&	3.60,  1.40		&	85.18	&	0.77		&	0.38		&	 20.98\\
sink 8, sink 17	&	13.34, 6.84		&	97.39	&	0.66		&	0.51		&	 65.26\\
sink 11, sink 28	&	2.40,  1.37		&	47.40	&	0.93		&	0.57		&	 53.89\\
sink 22, sink 30	&	6.47,  2.64		&	26.08	&	0.96		&	0.40		&	 43.84\\
sink 27, sink 25	&	2.44,  1.08		&	93.40	&	0.17		&	0.44		&	 81.49\\
sink 37, sink 29	&	5.75,  2.30		&	42.31	&	0.62		&	0.40		&	 62.75\\
sink 37, sink 38	&	5.75	,  2.42		&	30.59	&	0.32		&	0.42		&	 30.99\\
sink 40, sink 42	&	4.10,  1.64		&	31.47	&	0.95		&	0.40		&	 39.88\\
sink 44, sink 41	&	1.66,  1.45		&	60.57	&	0.85		&	0.87		&	104.2\\
sink 44, sink 46	&	1.66,  1.42		&	16.67	&	0.95		&	0.85		&	 29.14\\
sink 51, sink 46	&	1.48,  1.42		&	28.57	&	0.65		&	0.95		&	 23.47\\
sink 51, sink 55	&	1.48,  0.88		&	35.67	&	0.57		&	0.60		&	 15.32\\
sink 52, sink 53	&	3.05,  1.38		&	20.68	&	0.98		&	0.45		&	 32.91\\
sink 59, sink 61	&	1.54	,  1.49		&	18.85	&	0.94		&	0.97		&	 36.52\\
sink 61, sink 62	&	1.49,  8.61		&	35.38	&	0.86		&	0.57		&	 56.53\\
\hline

\end{tabular}
\end{table*}

\begin{table*} \label{tbl-4}
\centering
 \caption{Summary of the results for model M4. The columns include the indices of the sink particles which are components of binaries, the masses of the components, the semi-major axis (a), the eccentricity (e), the mass ratio (q), and the binary separation (d) at the end of the simulation. } 
 \begin{tabular} {ccccccc}
\hline
\hline
 Component indices & component masses ($M_{\odot}$) & a (au) & e &  q  &  d(au) \\
 \hline
sink 1, sink 3	&	14.26, 10.51		&	26.92	&	0.99		&	0.73		&	 51.39 \\
sink 3, sink 8	&	10.51, 5.25		&	38.82	&	0.92		&	0.50	    &	 73.28 \\
sink 5, sink 4	&	20.38, 17.47		&	18.24	&	0.14		&	0.85		&	 20.17\\
sink 6, sink 10	&	12.42, 7.95		&	26.87	&	0.20		&	0.64		&	 28.77\\
sink 9, sink 7	&	14.86, 12.79		&	17.22	&	0.90		&	0.86		&	 32.71\\
sink 16, sink 17	&	7.42,  6.80		&	6.10		&	0.95		&	0.92		&	 1.90\\
sink 17, sink 24	&	6.80, 3.68		&	69.58	&	0.92		&	0.54		&	 73.48\\
sink 22, sink 20	&	7.93,  1.70		&	19.20	&	0.89		&	0.21		&	 33.91\\
sink 22, sink 21	&	7.93,  6.58		&	27.75	&	0.96		&	0.82		&	 41.93\\
sink 25, sink 27	&	2.35,  1.59		&	32.01	&	0.06		&	0.67		&	 34.13\\
\hline

\end{tabular}
\end{table*}

\begin{table*} \label{tbl-5}
\centering
 \caption{Summary of the results for model M5. The columns include the indices of the sink particles which are components of binaries, the masses of the components, the semi-major axis (a), the eccentricity (e), the mass ratio (q), and the binary separation (d) at the end of the simulation.} 
 \begin{tabular} {ccccccc}
\hline
\hline
 Component indices & component masses ($M_{\odot}$) & a (au) & e &  q  &  d(au) \\
 \hline
sink 1, sink 10	&	19.80, 2.99		&	23.74	&	0.69		&	0.15		&	 40.11 \\
sink 2, sink 11	&	 9.46, 6.94		&	16.03	&	0.87		&	0.73	    &	 29.68 \\
sink 7, sink 77	&	6.17, 1.92		&	40.75	&	0.97		&	0.31		&	 55.02\\
sink 8, sink 13	&	26.34, 6.00		&	73.64	&	0.59		&	0.22		&	 34.26\\
\hline

\end{tabular}
\end{table*}

\begin{table*} \label{tbl-6}
\centering
 \caption{Summary of the results for model M6. The columns include the indices of the sink particles which are components of binaries, the masses of the components, the semi-major axis (a), the eccentricity (e), the mass ratio (q), and the binary separation (d) at the end of the simulation.} 
 \begin{tabular} {ccccccc}
\hline
\hline
 Component indices & component masses ($M_{\odot}$) & a (au) & e &  q  &  d(au) \\
 \hline
sink 1, sink 16	&	31.96, 5.49		&	37.02	&	0.57		&	0.17		&	 24.56 \\
sink 2, sink 10	&	17.57, 15.83		&	9.16		&	0.87		&	0.790   &	 16.70 \\
sink 4, sink 6	&	20.88, 4.36		&	85.68	&	0.65		&	0.20		&	 122.2\\
sink 8, sink 5	&	10.66, 9.52		&	18.80	&	0.99		&	0.89		&	 31.93\\
sink 10, sink 15	&	15.83, 8.68		&	57.23	&	0.92		&	0.54		&	 73.60\\
sink 24, sink 25	&	1.34,  0.88		&	18.19	&	0.97		&	0.66		&	 31.04\\
\hline

\end{tabular}
\end{table*}

\begin{table*} \label{tbl-7}
\centering
 \caption{Summary of the results for model M7. The columns include the indices of the sink particles which are components of binaries, the masses of the components, the semi-major axis (a), the eccentricity (e), the mass ratio (q), and the binary separation (d) at the end of the simulation. } 
 \begin{tabular} {ccccccc}
\hline
\hline
 Component indices & component masses ($M_{\odot}$) & a (au) & e &  q  &  d(au) \\
 \hline
sink 1, sink 4	&	67.50, 41.02		&	14.50	&	0.86		&	0.60		&	 25.27\\
sink 2, sink 5	&	48.54, 12.06		&	58.22	&	0.90		&	0.24		&	 110.0\\
sink 4, sink 3	&	41.02, 30.49		&	46.51	&	0.89		&	0.74		&	 53.54\\
\hline

\end{tabular}
\end{table*}

\subsection{Binary properties} 

\begin{figure*} 
    \centering
    \includegraphics[angle=0,scale=0.375]{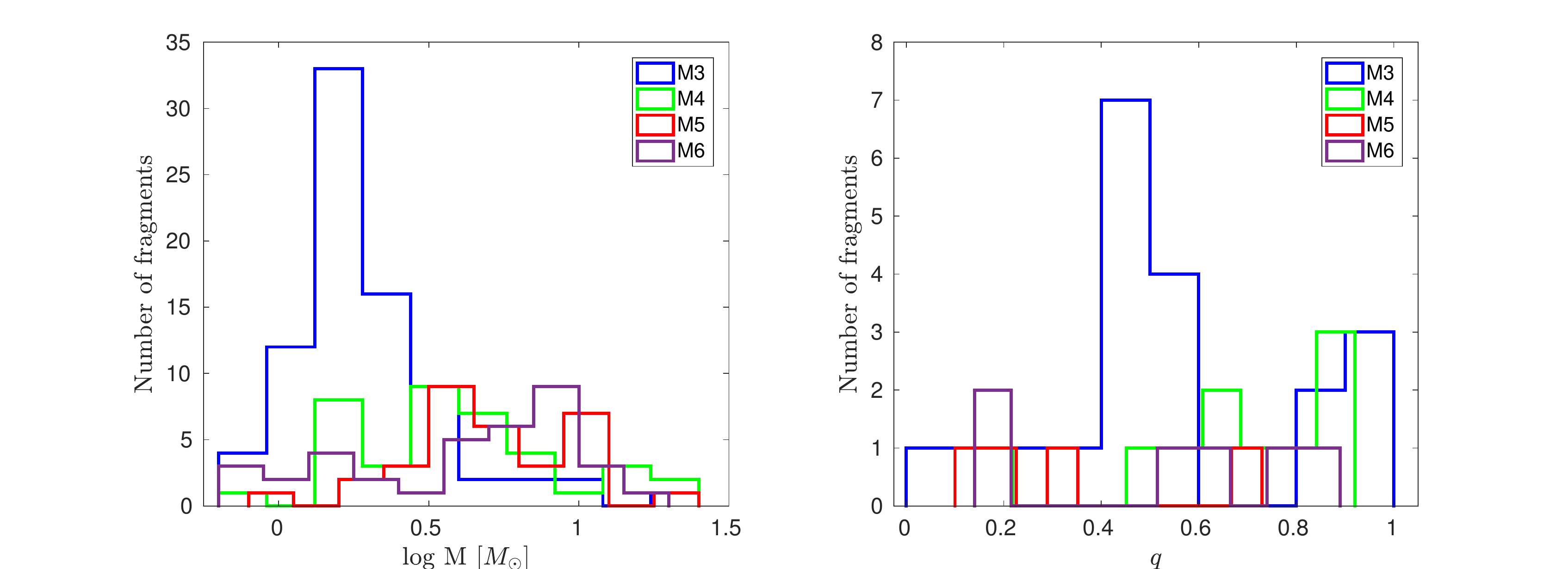}       
  \caption{Distribution of the masses of proto-binary components (on the left) and the distribution of the binary mass ratio (on the right) which have been formed in models M3 - M6 with initial rotation. In each part, the top left panel is for model M3, the bottom left panel is for model M4, the top right panel is for model M5, and the bottom right panel is for model M6. The masses are given in units of solar mass $M_{\odot}$. The histograms are computed when a total of 200 $M_{\odot}$ is contained in all of the  protostars formed in each model. Colour in online edition.}\label{fig:13}
     \end{figure*}

\begin{figure*}
    \centering
   \includegraphics[angle=0,scale=0.375]{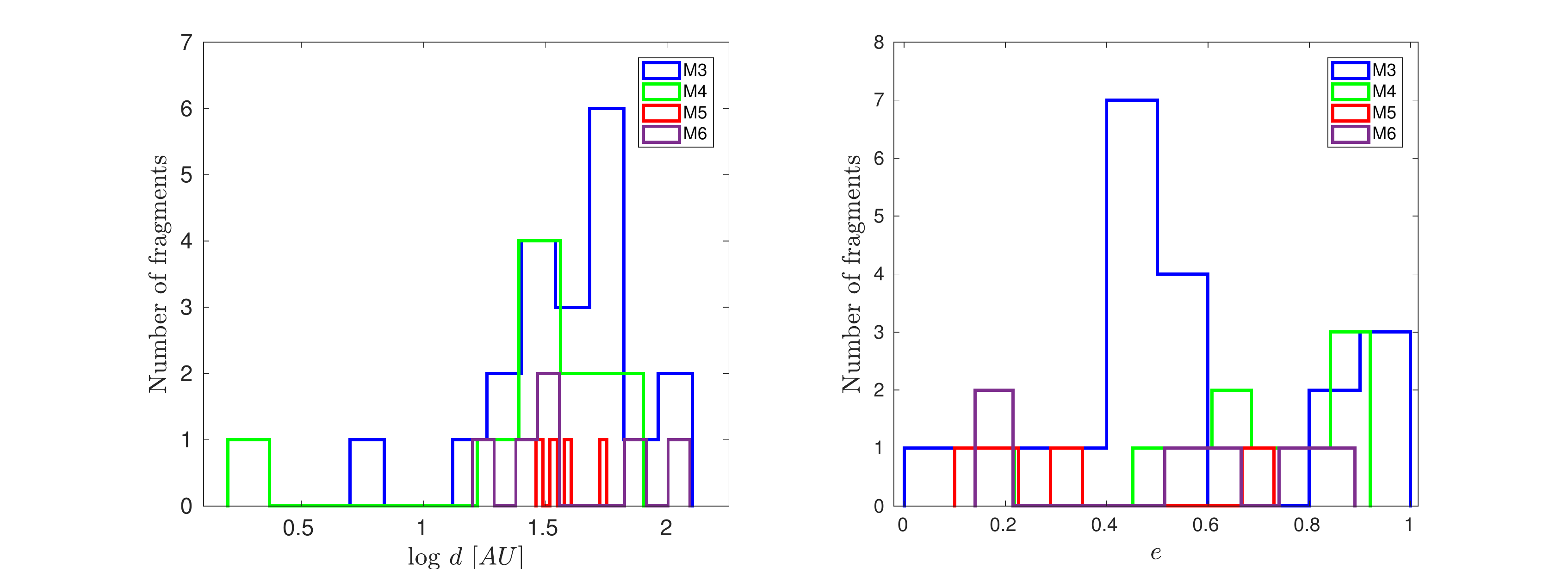}     

  \caption{Distribution of the binary separation (on the left) and the distribution of binary eccentricity (on the right) which have been formed in models M3 - M6 with initial rotation. In each part, the top left panel is for model M3, the bottom left panel is for model M4, the top right panel is for model M5, and the bottom right panel is for model M6. The masses are given in units of solar mass $M_{\odot}$. The histograms are computed when a total of 200 $M_{\odot}$ is contained in all of the protostars formed in each model. Colour in online edition.} \label{fig:14}
     \end{figure*}

We finally turn to a detailed investigation of the properties of the protobinaries which are formed in our smimulations. Figures \ref{fig:13} and \ref{fig:14} show histograms of the masses of the proto-binary components, as well as the binary mass ratio, the binary separation and the binary eccentricity. 
The histograms in the left part of Figure \ref{fig:13} show the distribution of the mass of the proto-binary components in models M3, M4, M5, M6, respectively. To explore how the process of fragmentation depends on the turbulent Mach number $\mathcal{M}$ and the rotational parameter $\beta{}$, we divide the four models into two sets (1 \& 2) and four subsets (1.1, 1.2, 2.1, and 2.2). Set 1.1 contains models M3 \& M4 with Mach number $\mathcal{M}$ varying from $0.5$ to $1.0$, while keeping the rotational parameter constant at the low value $\beta{} = 5\%$. Set 1.2 consists of models M6 \& M5 with Mach number $\mathcal{M}$ varying from $0.5$ to $1.0$ while keeping the rotational parameter constant at the higher value $\beta{}= 10\%$. We observe that a primordial gas cloud with more rotational support and relatively strong transonic turbulence produces higher mass protostars. For instance, model M4 and M5 both are transonic but model M5 with stronger rotational support produces more higher mass protostars than model M4. Similarly, set 2.1 contains model (M3 \& M6) with rotational parameter $\beta{}$ varying from $5\%$ to $10\%$ while keeping $\mathcal{M}$ constant at 0.5. Set 2.2 contains models M4 \& M5 where the rotational parameter $\beta{}$ varies from $5\%$ to $10\%$ while keeping $\mathcal{M}$ constant at 1.0. Comparing the histograms suggests that the number of Pop.~III protostars and their individual masses are controlled by the strength of the initial rate of rotation $\omega{}$ as well as the Mach number of the turbulence $\mathcal{M}$ in the primordial gas clouds.  

We now focus on the properties of the Pop.~III protostellar systems in models M3 - M6. There are a total of 40 Pop.~III protostellar binaries in the rotating gas cloud models at the end of the simulations. The number of binaries in each model is summarized in table~1. Furthermore, in table~2 we provide the total mass contained within these binary systems as well as the total mass of the isolated Pop.~III protostars. With respect to the total mass of $200$~M$_{\odot}$ which is required to be inside sink particles at the end of the simulations, we also calculated the fraction of this total mass which is part of Pop.~III binaries. We find that model M4 ($\mathcal{M} = 1.0$ \& $\beta{} = 5\%$) yields 10 proto-binary systems, while model M3 ($\mathcal{M} = 0.5$ \& $\beta{} = 5\%$) produces $20$ proto-binary systems. Despite the smaller number of binaries formed in model M4 it seems that the stronger turbulence in this model leads to $10\%$ more mass contained within binaries compared to model M3. In model M5 ($\mathcal{M} = 1.0$ \& $\beta{} = 10\%$), for comparison, we observe only 4 binaries, possibly indicating that rotation plays a role in suppressing the formation of proto-binary systems. 
\citet{b43} have analyzed the mass distribution of Pop.~III binary systems. They report an upper limit for the most massive sink which is around $40$~M$_{\odot}$. The mass range in our rotating models M3 - M6 goes from  $31.96$~M$_{\odot}$ and $0.88$~M$_{\odot}$, respectively. However, if we include the Pop.~III protobinaries formed in the non-rotating model M7, then mass of the most massive proto-binary component is found to be $67.50$~M$_{\odot}$.

In the right part of Figure \ref{fig:13}, we present the distribution of the binary mass ratio $q$ in the rotating models M3 - M6. A detailed summary of the results can be found in tables 4, 5, 6, and 7. The top left, bottom left, top right and bottom right panels show the distribution of the mass ratio q for models M3, M4, M5, and M6; respectively. To analyze how the distribution of q has responded to the varying parameters ($\mathcal{M}$ \& $\beta{}$), we use the same comparison scheme as for the left part of the Figure.  
Comparing models (M3 \& M4) and (M5 \& M6) reveals that Pop III proto-binary systems with higher mass ratios can be formed in collapsing primordial gas clouds with subsonic turbulence levels. Similarly, by comparing models (M3 \& M6) and (M4 \& M5), we find that a weak initial rotational support seems to increase the number of binary systems with a high mass ratio q. In general, a wide spectrum of mass ratio $q$ within the range 0.009 - 0.97 has been observed in all of our rotating models. We also see that with a decreasing turbulence level and rotational support the range of mass ratio q gets seems to get widened in our simulations.
The broad overall distribution of mass ratio q in our Pop.~III protobinaries agrees well with \citet{b43} who have also found a wide spectrum for the mass ratio $q$. Here it is interesting to note that they include mergers while we do not.

In figure \ref{fig:14} we present on the left the distribution of binary separations $d$ and on the right the distribution of binary eccentricity $e$ in the rotating cloud models M3 - M6. The histograms on the left show the distribution of the binary separations $d$ in the top-left, bottom-left, top-right, and bottom-right panels for models M3, M4, M5, and M6, respectively. 
Comparing the sets (M3 \& M4) and (M5 \& M6) allows us to address the effect of the turbulent Mach number on these distributions. It is evident that under weak rotational support both subsonically and transonically turbulent gas clouds allow for the formation of relatively tightly bound ($6.02$~au \& $1.90$~au) proto-binary systems. However, the upper end of the binary separation distribution seems to be limited to around $\lesssim75$~au in the case of transonic turbulence whereas the subsonically turbulent models with identical rotational support yield protobinaries with separations as large as $116.5$~au. For the case of strong rotational support the cloud models, however, do not permit the formation of tightly bound proto-binary systems and the lower end of the binary separation distribution increases to $29.68$~au \& $16.70$~au, respectively. Also a significant decrease in the upper limit of the binary separation distribution is observed in the case of the transonically turbulent gas cloud where the maximum value of the binary separation to $55.02$~au in contrast to $122.2$~au in the subsonic case.

Comparing model sets (M3 \& M6) and (M4 \& M5) allows us to investigate the influence of the initial rotational support on subsonically and transonically turbulent gas clouds, respectively. For the models with subsonic turbulence (M3 \& M6), the initial rotational support does not affect much the upper end of around $120$~au of the binary separation distribution. However, only in the case of weakly rotationally supported gas clouds with subsonic turbulence the lower end is populated with more tightly bound Pop.~III proto-binary systems with binary separations as small as $6.02$~au. Similarly, in the models with transonic turbulence (M4 \& M5) we observe that the rotational support repeats its role of populating the lower end of the distribution of binary separation. Weak initial rotational support produces tightly bound proto-binary systems with a binary separation as small as $1.90$~au.  
  
Turning our attention to the right side of Figure \ref{fig:14}, we show the distribution of eccentricity $e$ obtained in models (M3 - M6). A summary can also be found in tables 4 - 7. The top left, bottom left, top right, and bottom right panels show the distribution of $e$ for models M3, M4, M5, and M6, respectively. Again we compare the model sets we discussed before.
To see how turbulence has affected the distribution of the binary eccentricities, we compare models (M3 \& M4) and (M5 \& M6). The eccentricities of the Pop.~III proto-binary systems are evidently strongly affected by the initial turbulent level in the clouds. We find that the eccentricities for all binary systems formed in the models are distributed within the range 0.14 - 0.99. Subsonic initial turbulence ($\mathcal{M}=0.5$) in the parent gas cloud yields more eccentric orbits for the emerging proto-binary systems. Similarly, comparison of  models (M3 \& M6) and (M4 \& M5) identifies a general trend in the distribution of the binary eccentricities. It is evident that primordial gas with weaker initial rotational support $\beta{}=0.05$ produces protobinaries with more eccentric orbits.

\subsection{Convergence tests based on varying $N_{opt}$ and $M_{total}$} 

In this section, we aim to assess the main uncertainties in our simulations which are related to  the number of neighbours $N_{opt}$ used in the SPH scheme, as well as the mass $M_{total}$ which must be accreted by the protostars before the simulations are terminated.

The quantity $N_{opt}$ only equals the true number of neighbours as long as the number density of particles within the smoothing sphere is approximately constant. So at best the $N_{opt}$ parameter characterizes the mean neighbour number – and there can be strong fluctuations about this mean, for example in strong density gradients (see for example \citealt{Price2012}).

We present a set of convergence tests to validate the results reported in this work. These tests report on the binary statistics obtained from five models, M3A, M3B, M3C, M3D, and M3E with number of neighbours $N_{opt}$ = 50, 75, 100, 125, and 150 and $M_{total}$ = 50$~M_{\odot}$. We further analyze 6 evolutionary stages of model M3 when the total mass $M_{total}$ inside protostars has reached values of 50$~M_{\odot}$, 100$~M_{\odot}$, 200$~M_{\odot}$, 300$~M_{\odot}$, 400$~M_{\odot}$, and 500$~M_{\odot}$, respectively. These evolutionary stages are labeled as M3F, M3G, M3H, M3I, M3J, and M3K. In this run $N_{opt}$ is set to our standard value of 50. The results obtained from these models are listed in tables 8 and 9. 
 
In Figure 15, we show the spatial distribution of protostars which have formed at the stages M3F-M3K of the model M3. The 6 plots from top-left to the bottom-right panel show the stage when the total mass contained within sinks has reached values from 50$~M_{\odot}$ to 500$~M_{\odot}$, respectively. The last two stages of the collapse show the ejection of some of the protostars from the cluster by three- and four-body dynamical encounters in dense stellar systems \citep{Gvaramadze01, Gvaramadze02, Pflamm, Kroupa98}.

In the top-left panel of Figure 16, we show the mass distribution of all sinks when the evolution of models M3A-M3E has reached the stage when the total mass of the sinks is $M_{total}$ = 50$~M_{\odot}$. The figure suggests that despite changing the number of neighbours $N_{opt}$, the mass distribution of the sinks remains approximately the same. For all values of $N_{opt}$, the mass of the majority of the sinks falls within the range of 0.650$~M_{\odot}$ - 5.625$~M_{\odot}$.

The top-right panel shows the mass distribution of the sinks for the series of model stages M3F-M3K. 
We explore values for the total mass $M_{total}$ of the sinks ranging from 50$~M_{\odot}$ - 500$~M_{\odot}$. In this case the mass distribution also remains more or less identical even when the evolution of the models reaches various stages. The shape of the mass distribution indicates that most of the sinks have masses within a range of 0.650$~M_{\odot}$ - 10.0$~M_{\odot}$. However, the total number of sinks which appear during the collapse increases with time (see table 9). When $M_{total}$ has reached 500$~M_{\odot}$, the total number of sinks has climbed to 144.

The bottom-left panel of Figure 16 provides a comparison of the mass ratio $q$ for all simulation models which vary the number of neighbours $N_{opt}$. The mass ratio distribution for models M3A-M3E shows that on average, varying $N_{opt}$ always leads to a similar distribution for $q$ which covers the entire range of possible values. The bottom-right panel of  Figure 16 represents the mass ratio distribution of the protobinaries for the model stages M3F-M3K. It is evident that for the different evolution stages with $M_{total}$ going from 50 $~M_{\odot}$ to 500 $~M_{\odot}$, there appears to be no bias for the mass ratio $q$ and on average its distribution covers the entire range from 0 to 1.

\begin{table}
\caption{Summary of the values for $N_{opt}$ and the final outcome of the simulation models M3A-M3E. The initial conditions are identical to model M3 for each model. The simulations are terminated when a total of 50$~M_{\odot}$ has been accreted by the protostars.}             
\label{table:8}      
\centering                          
\begin{tabular}{c c c c c c}        
\hline\hline                 
Model & \# of $N_{opt}$  & \# of protostars & \# of binaries \\    
\hline                        
M3A  & 50     & 12  &  4\\
M3B  & 75    & 14  &  5\\
M3C  & 100	  & 12  & 3\\
M3D  & 125    &  9  & 2\\
M3E  & 150    &  9  & 2\\   
\hline
\end{tabular}
\end{table}

\begin{table}
\caption{Summary of the total mass inside protostars $M_{total}$, the total number of protostars and the number of binary systems at the evolutionary stages M3F-M3K for model M3. The initial conditions are identical to models M3A-M3E.}             
\label{table:9}      
\centering                          
\begin{tabular}{c c c c c c}        
\hline\hline                 
Model & $M_{total}$($~M_{\odot}$)  & \# of protostars & \# of binaries \\
\hline                        
M3F  & 50      & 21  &  9\\
M3G  & 100     & 46  &  14\\
M3H  & 200     & 81  & 20\\ 
M3I  & 300	   & 108 & 24\\
M3J  & 400     & 118 & 20\\
M3K  & 500     & 144 & 14\\
\hline
\end{tabular}
\end{table}	

Looking further at the binary properties, Figure 17 presents distributions of the semi-major axis $a$ and eccentricity $e$ for the protobinaries in the set of models M3A-M3E and the simulation stages M3F-M3K. The top-left panel indicates the distribution of the semi-major axis obtained for the set M3A-M3E. As discussed above, this set of models varies the number of neighbours $N_{opt}$ in the range from 50 - 150. Apart from the extreme cases of $N_{opt}$ = 125 and 150, the models in this set seem to produce a wide range of semi-major axes $a$ (17 AU - 95 AU). 
The top-right panel shows the semi-major axis distributions for the model stages M3F-M3K. Interestingly, we observe a trend for more proto-binary systems to appear with relatively larger semi-major axes $a$ with progressively increasing $M_{total}$. However, for the extreme case of $M_{total}$ = 500$~M_{\odot}$ (the last evolutionary stage of model M3 considered), we find that the number of proto-binary systems with large values of $a$ has sharply declined. 
We believe that three-body interactions (see for example \citet{Susa2014}) which start to become dominant from the evolution stage of $M_{total}$ = 400$~M_{\odot}$ are the prime cause of disruption of some of the widely separated proto-binary systems. Such dynamical encounters can affect the number of proto-binary systems in a cluster. This dynamical effect is illustrated in the last two panels of figure 15 in which ejection of protostars can be seen. 
The bottom left panel of Figure 17 illustrates how the distribution of eccentricity $e$ of the Pop.~III proto-binary systems responds to variations in number of neighbour $N_{opt}$. Just like the mass ratio $q$, the distribution of eccentricity $e$ covers the entire range of 0 to 1. Despite the reduced number of 9 fragments in the extreme case of $N_{opt}$ = 150 and hence the small number of only 2 protobinaries (see table 8), both high and low eccentric orbits seem to appear alongside each other. The bottom-right panel describes the distribution of eccentricity $e$ for the model stages M3F-M3K. We observe a general trend of high eccentric orbits for Pop.~III proto-binary systems which form at various stages of primordial gas cloud collapse. The overall distribution does not change appreciably when $M_{total}$ increases during the cloud collapse. 
A final important matter is the conservation of angular momentum in simulations that produce a lot of sink particles. We have checked the total combined orbital angular momentum of the sink particles and the SPH particles over the course of the long simulation M3, and find that, despite the fact that this calculation does not take the angular momentum accreted by the sink particles into account, the angular momentum conservation is good to within 0.001 $\%$.







  \begin{figure*}
\centering
\includegraphics[angle=0,scale=0.34]{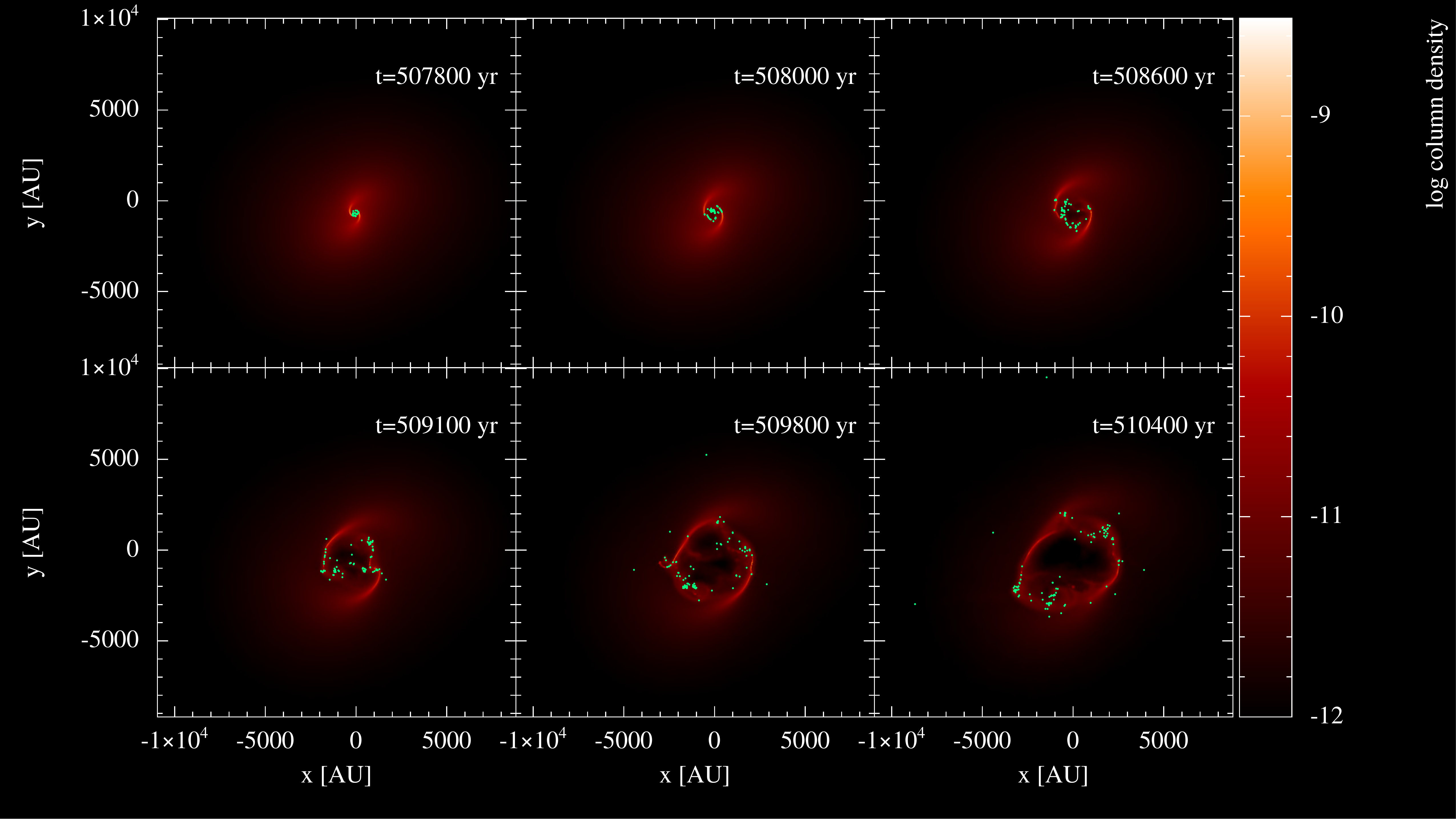}
\caption{From top-left (first panel) to bottom-right (sixth panel), results are shown for the simulation stages M3F-M3J for model M3. The plots show face-on views of the column density for the total gas density integrated along the rotation axis(z). The colour bars on the right of each panel show the logarithm of the column density ($\Sigma{}$) in physical units of g cm$^{-2}$. Each calculation was performed with 1150709 SPH particles. Colour in online edition. }\label{fig:15}
\end{figure*}


  

   \begin{figure}
   \includegraphics[width=9.650cm]{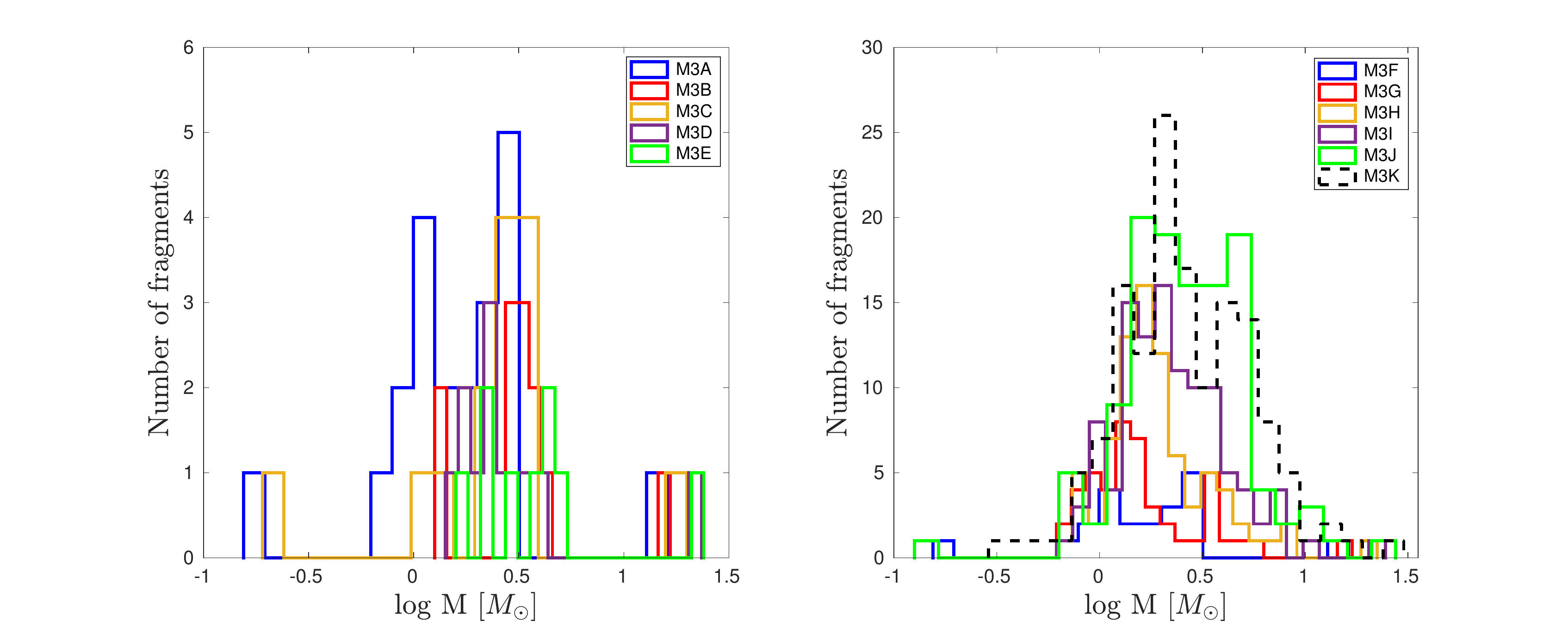}
   \includegraphics[width=9.650cm]{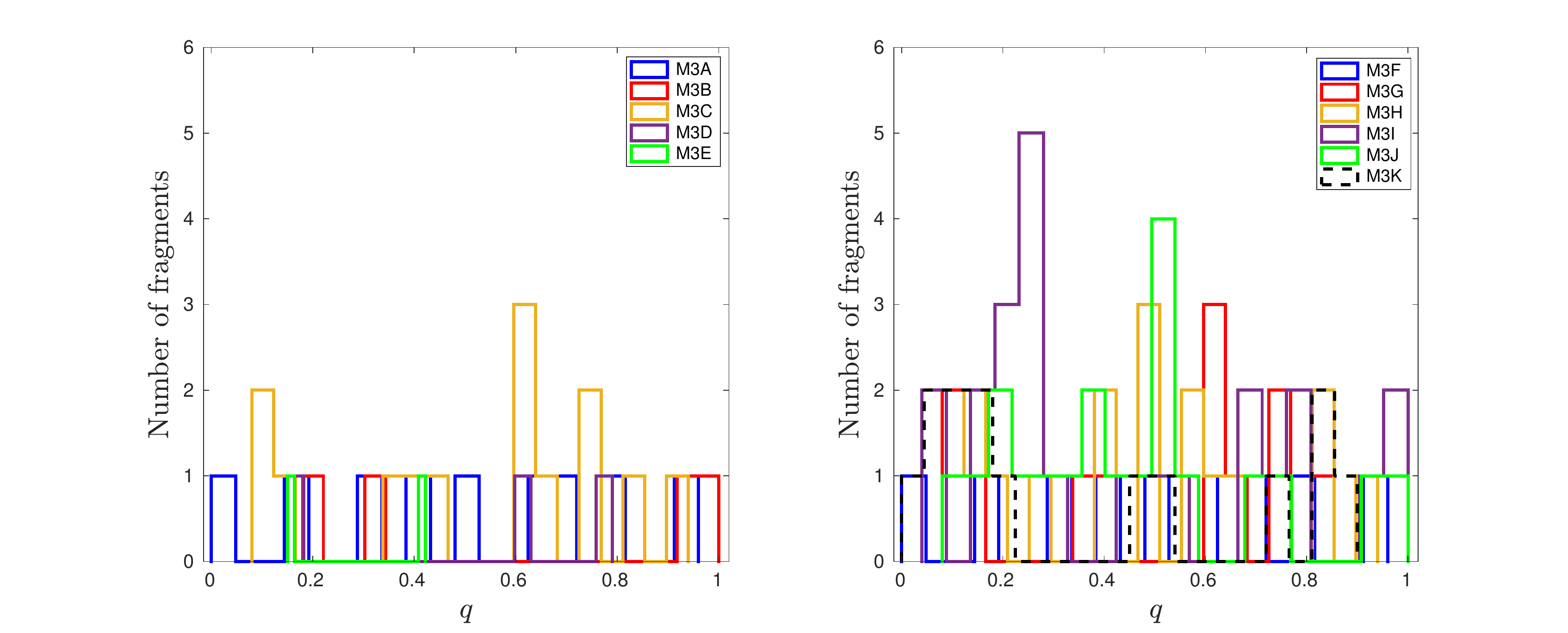}
      \caption{Distributions of the masses of proto-binary components for models M3A-M3E (top left) and the simulation stages M3F-M3K for model M3(top right). For the same set of models, the distributions of the binary mass ratio are shown in the bottom left and bottom right plots, respectively. Colour in online edition. }\label{fig:16}
   \end{figure}


   \begin{figure}
   \includegraphics[width=9.650cm]{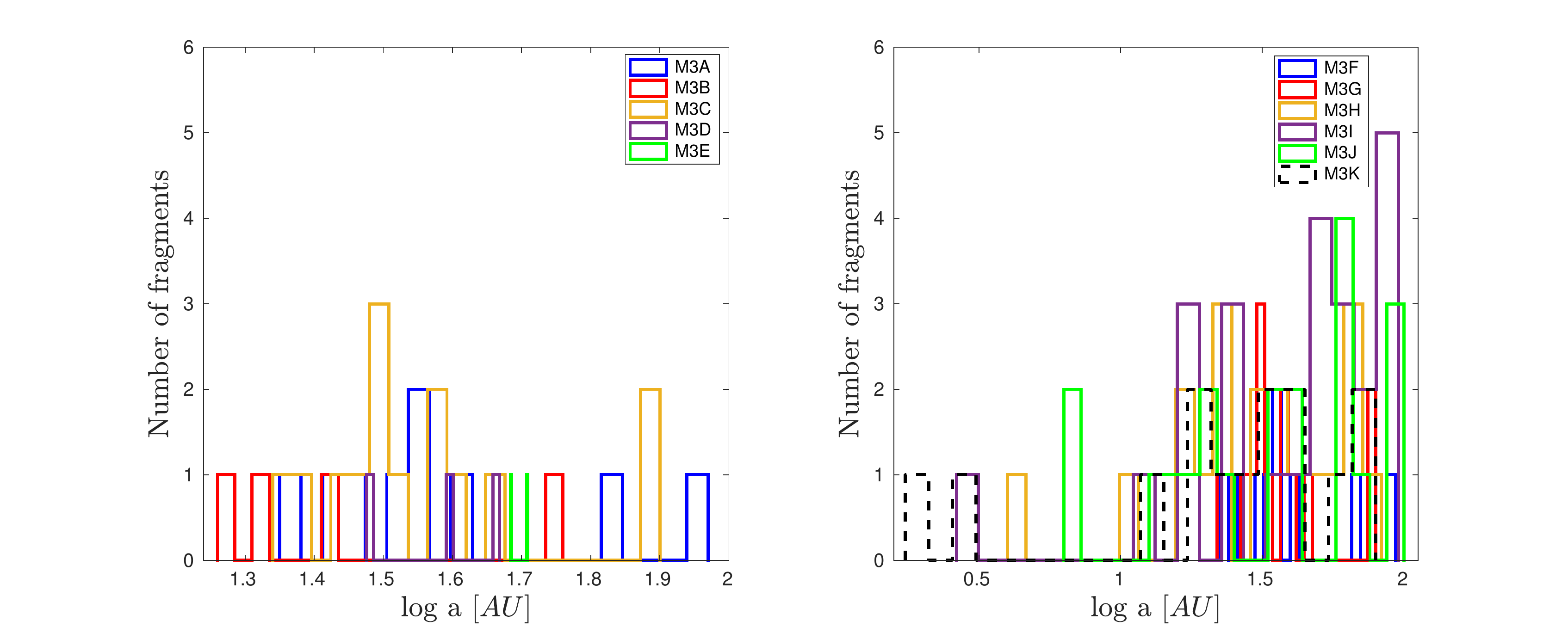}
   \includegraphics[width=9.650cm]{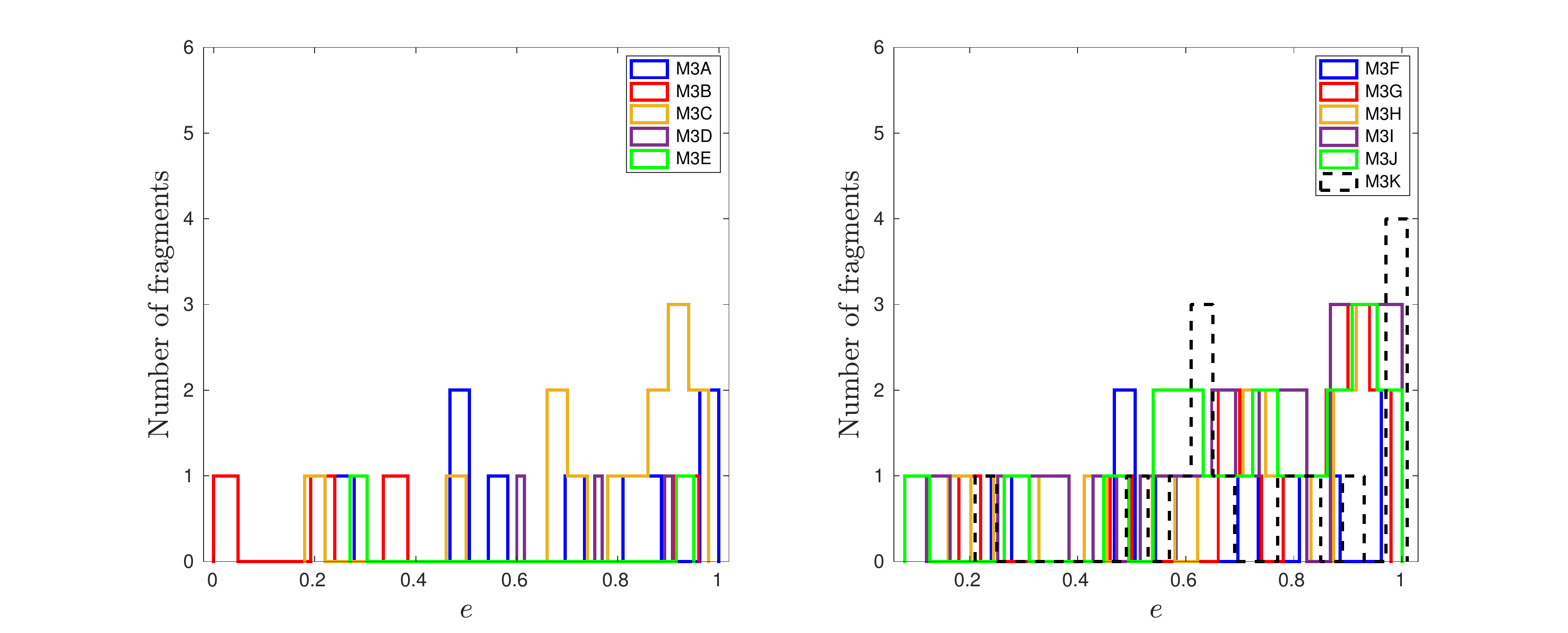}
      \caption{Distributions of the semi-major axis of protobinaries for models M3A-M3E (top left) and simulation stages M3F-M3K for model M3(top right). For the same set of models, the distributions of the eccentricity are shown in the bottom left and bottom right plots, respectively. Colour in online edition. }\label{fig:17}
   \end{figure}

\subsection{Limitations of the present study and future work}

Our work is an attempt to study the possibility of Pop.~III protostar formation in an environment which is independent of both the radiation feedback from Pop.~III stars affecting nearby fragments in the collapsing primordial gas and the external effect of a UV radiation background. Although we consider three types of turbulence (subsonic, transonic, and supersonic turbulence) to investigate their effects on the Pop.~III star formation, a  wider range of initial supersonic turbulent levels should be explored as supersonic turbulence, in particular, plays an important role in shaping the cloud structure, and in controlling star formation, because it creates the seeds for local gravitational collapse \citep{Elmegreen04, Mac Low04, McKee07}. 


In the present work we assume that the star formation process is confined to an individual isolated cloud of primordial gas. This Jeans unstable cloud is chemically evolved from an initial metal-free gas state to yield the first generation of protostars. However, since massive stars may eventually explode as Pop.~III supernovae (SNe) and seed the primordial host cloud with heavy elements \citep{Tsujimoto99, Wise2011}, this metal enrichment can have strong implications for the formation of subsequent generation of stars, due to fragmentation triggered by dust or metal line cooling  \citep{Ricotti02, Ciardi05, Ricotti08, Wise08, Smith09, bovino2014, Pallottini2014, Bovino2016}. The chemical feedback and enrichment processes can also affect the process of gas accretion in evolving Pop.~III proto-binary systems and can even alter their binary properties as the composition of the cloud changes. While we do not investigate these processes in this paper, it is clear that they can play a major role in the subsequent evolution and the transition to the formation of low-mass stars. A possible approach to study the related mixing processes via SPH simulations has been outlined by \citet{Greif10}.

The reported work in this paper has another limitation as we currently do not include mergers between protostars within our simulations. \cite{b43} estimate that up to 50 $\%$ of their sink particles are lost to mergers. Since mergers can therefore strongly affect the mass spectrum of the fragments which has a direct impact on the Pop.~III proto-binary population, the absence of sink mergers makes our work different from the previous investigations reported by \citet{b24, Stacy2011, b18, Stacy2012} in which mergers of sinks have been allowed. Also, it has been pointed out by \citet{b18} that the details of merging algorithm and possible modifications within the scheme can have profound effects on the sink accretion history. As a result of the impact of binary mergers, the Pop.~III proto-binary population could therefore be reduced in size with binary properties that are modified compared to what has been reported in this work.
On the other hand, the size of the protostars is very much smaller than the accretion radius and the critical distance for merging adopted by \citet{b43} is of the same order of magnitude as the accretion radius. We therefore assume that they may have overestimated the rate of mergers and the results obtained here may be more realistic.

In addition, magnetic fields could alter the dynamics in the primordial gas. Even in case of initially weak magnetic fields, efficient amplification may take place via the small-scale dynamo \citep{Latif2014, Schober12, Schleicher10, Sur10}. After the formation of an accretion disk, rotation may order the magnetic fields and start driving jets and outflows \citep{Latif2016, Machida13}. Further amplification may occur in subsequent supernovae and HII regions, (see for example \citealt{Seifried2014, Koh2016}).

\section{Conclusions and outlook}

We have conducted numerical simulations to study the fragmentation process of primordial gas clouds with our new code GRADSPH-KROME, with the main goal to assess the formation of protostellar binaries and their key properties like the typical binary separation and the distribution of their masses. The code combines the SPH framework with the chemistry package KROME, which allows us to evolve the models keeping track of a detailed chemical reaction network that determines the gas cooling and is of crucial importance to be able to follow the details of the fragmentation process. We have validated the GRADSPH-KROME code presented here, showing that the well-known thermal evolution of primordial gas clouds can be reproduced under spherically symmetric conditions. The simulations were evolved until a total of $200$~M$_\odot$ has been accreted into sinks, so that the output can be compared at the same evolutionary stage of the collapse. 

In non-rotating and subsonically as well as transonically turbulent gas clouds, we observed a single centrally located fragment, which accretes material with a mean accretion rate of $6.345 \times 10^{-1}$~M$_{\odot}$yr$^{-1}$ from its surrounding  gas and reaches $200$~M$_{\odot}$ when we terminate the simulation. In the presence of supersonic turbulence, however, fragmentation occurs and produces $11$ fragments at various spatial locations even in the absence of an initial net rotation. Some of these fragments become proto-binary systems, accrete material from the surrounding gas and show pronounced variability in their mass accretion history. We classified these protostars into the most actively accreting and the least actively accreting proto-binary components in terms of their mass accretion rates. Unlike in the other non-rotating clouds including subsonic/transonic turbulence, the supersonically turbulent model provides a different environment for the accretion process. The mean mass accretion rate for the most active and for the least active proto-binary components correspond to $1.896 \times 10^{-1}$~M$_{\odot}$yr$^{-1}$ and $8.804 \times 10^{-2}$~M$_{\odot}$yr$^{-1}$, respectively.

Strong filamentary structures are key features of all turbulent gas models with an initial solid body rotation. While the ejection of lighter mass fragments cloud happen in small clusters of Pop.~III protostars \citep{b43}, we have not found such events within the comparatively early evolution time of the models explored here. In rotating clouds all protostars tend to form in dense filamentary structures. Due to rotation and the gravitational dynamics between protostars, they are subsequently distributed towards regions which are mostly not related to any filamentary structure. The overall presence of central disks in our models shows that gravitational disk instabilities are important for regulating the process of fragmentation.

In order to compare the mass accretion rates between the rotating primordial gas cloud models, we again distinguish the most active and the least active protostars in terms of their mass accretion. Our rotating gas models have shown various levels of fragmentation primarily dependent on the initial conditions in these clouds. We have observed a total of 40 Pop.~III proto-binary systems. Based on our classification we have found  maximum and minimum accretion rates of $2.31 \times 10^{-1}$~M$_{\odot}$yr$^{-1}$ and $2.18 \times 10^{-4}$~M$_{\odot}$yr$^{-1}$, respectively.

The mass spectrum of the individual Pop.~III proto-binary components covers a range from $0.88$~M$_{\odot}$ to $31.96$~M$_{\odot}$ and is found to sensitively depend on the Mach number $\mathcal{M}$ as well as the rotational parameter $\beta{}$ of the clouds. Based on our simulations the mass ratio is found to vary from $0.009$ to $0.97$ for all of the proto-binary systems. For decreasing $\mathcal{M}$ and $\beta{}$ our results suggest that proto-binaries with high mass ratio are more easily formed. The eccentricity is reported to be strongly dependent on the initial conditions. We have observed eccentricities ranging from $0.14$ to $0.99$.  The distribution of binary separation is also obtained in this work, and was found to vary between separations as low as $1.90$~au up to values of $\sim120$~au.

While the specific results obtained here show some dependence on the initial conditions, the formation of binary systems appears as a generic phenomenon at least at this early evolutionary stage of the system. For binaries with separations of order $\sim100$~au, it seems plausible that they will survive for a long time and could become permanently stable, in particular when the gas from the envelope is depleted. We suggest that the preliminary trends obtained in this work should be corroborated using a set of simulations exploring a larger range of initial conditions. It is clearly necessary to assess their longer term evolution in future work and also include the impact of mergers on the evolution of the primordial binary population in order to shed light on interesting phenomena such as the formation of X-ray binaries and possible gravitational wave events related to merging binary systems which may be observable with LIGO in the near future.

\section{Acknowledgements}
This research has made use of the high performance computing clusters Geryon2 and Leftraru. The first author RR gratefully acknowledges support from the Department of Astronomy of the University of Concepcion, Chile. The first author RR and the fourth author DRGS thank for funding through the Concurso Proyectos Internacionales de Investigaci\'on, Convocatoria 2015" (project code PII20150171). DRGS further thanks for funding via Fondecyt regular (project code 1161247) and via the Chilean BASAL Centro de Excelencia en Astrof\'isica yTecnolog\'ias Afines (CATA) grant PFB-06/2007, and ALMA-Conicyt project (proyecto code 31160001) via Quimal (project number QUIMAL170001), and via the Anillo (project number ACT172033). The second author SB thanks for funding through the DFG priority program “The Physics of the Interstellar Medium” (projects BO 4113/1-2). The third author SV developed the computer code which has been used in this work on the HPC system Thinking at KU Leuven which is part of the infrastructure of the FSC (Flemish Supercomputing Center). He thanks the KU leuven support team for helping with the use of this system. The third author also gratefully acknowledges the support of Prof. Dr. Stefaan Poedts and Prof. Dr. Rony Keppens for having provided both access and funding which made the use of the KU Leuven HPC infrastructure possible. He also thanks prof. Dominik Schleicher to provide access to the HPC cluster Leftraru. 


\label{lastpage}

\end{document}